\documentclass[12pt]{elsarticle}
\newcommand{\eref}[1]{(\ref{#1})} 
\usepackage[utf8]{inputenc}
\usepackage{graphicx}
\usepackage{color}
\usepackage{epstopdf}
\usepackage{booktabs}
\usepackage{pdflscape}

\expandafter\let\csname equation*\endcsname\relax
\expandafter\let\csname endequation*\endcsname\relax
\usepackage{amsmath}

\newcommand{\Ham}{\mathcal{H}}     
\newcommand{\spin}[1]{\sigma_{#1}} 
\newcommand{\nn}[2]{\left<#1,#2\right>} 
\newcommand{\nnn}[2]{\left<\left<#1,#2\right>\right>} 
\newcommand{\plaq}[4]{\left[#1,#2,#3,#4\right]} 
\newcommand{\nnsum}{\sum\limits_{\nn{i}{j}}\spin{i}\spin{j}} 
\newcommand{\nnnsum}{\sum\limits_{\nnn{i}{j}}\spin{i}\spin{j}} 
\newcommand{\plaqsum}{\sum\limits_{\plaq{i}{j}{k}{l}}%
  \spin{i}\spin{j}\spin{k}\spin{l}}

\newcommand{\tablesize}{\scriptsize} 

\newcommand{\betainfall}{0.551\,334(8)}

\usepackage{colortbl}
\definecolor{lightgrey}{rgb}{0.9,0.9,0.9}
\definecolor{grey}{rgb}{0.7,0.7,0.7}

\newcommand{\hllg}{\cellcolor{lightgrey}} 
\newcommand{\hlg}{\cellcolor{grey}} 

\begin{document}

\title
{ Multicanonical analysis of the plaquette-only gonihedric Ising model and its dual}

\author[itp]{Marco Mueller}
\ead{Marco.Mueller@itp.uni-leipzig.de}
\author[hwu]{Desmond A. Johnston}
\ead{D.A.Johnston@hw.ac.uk}
\author[itp]{Wolfhard Janke}
\ead{Wolfhard.Janke@itp.uni-leipzig.de}
\address[itp]{Institut f\"ur Theoretische Physik, Universit\"at Leipzig,\\ Postfach 100\,920, D-04009 Leipzig, Germany}
\address[hwu]{Department\ of Mathematics and the Maxwell Institute for Mathematical Sciences, Heriot-Watt University, Riccarton, Edinburgh, EH14 4AS, Scotland}


\begin{abstract}
  The three-dimensional purely plaquette gonihedric Ising model and its dual 
  are investigated to resolve inconsistencies in the literature for the 
  values of the inverse transition temperature of the very strong 
  temperature-driven first-order phase transition that is apparent in the 
  system. 
%
  Multicanonical simulations of this model allow us to measure system
  configurations that are suppressed by more than 60 orders of magnitude
  compared to probable states. With the resulting high-precision data, we find
  excellent agreement with our recently proposed nonstandard finite-size
  scaling laws for models with a macroscopic degeneracy of the low-temperature
  phase by challenging the prefactors numerically. We find an overall
  consistent inverse transition temperature of 
    $\beta^\infty = \betainfall$ 
  from the simulations of the original model both with periodic and fixed boundary
  conditions, and the dual model with periodic boundary conditions. For the original
  model with periodic boundary conditions, we {obtain} the first reliable estimate
  {of} the interface tension
  {$\sigma = 0.12037(18)$},
  using the statistics {of}  suppressed configurations.
\end{abstract} 


\maketitle

\section{Introduction}

The gonihedric Ising model was originally formulated by Ambartzumian and 
Savvidy~\cite{discretization} as a possible lattice discretization of
{an alternative ``linear'' action for}  the string worldsheet in bosonic string theory. 
{These} early discretizations {using} triangulations were then translated {to
plaquette surfaces generated as the spin cluster boundaries 
of classical generalized Ising spin Hamiltonians} by Savvidy and Wegner~\cite{spinssystem}.
For a recent review, see~\cite{Johnston2008}.

The {resulting} gonihedric Ising model is a generalized three-dimensional Ising model, where spins
$\spin{ }$, living on a three-dimensional cubic lattice, interact via
nearest $\nn{i}{j}$, next-to-nearest $\nnn{i}{j}$ and plaquette interactions
$\plaq{i}{j}{k}{l}$ and the weights of the different interactions are fine-tuned
so that the \emph{area} of spin cluster boundaries does not contribute to the 
partition function, but {rather} their edges and self-interactions. This leads to the family of Hamiltonians 
\begin{equation}
  \Ham^\kappa = -2\kappa\nnsum+\frac{\kappa}{2}\nnnsum-\frac{1-\kappa}{2}\plaqsum,
  \label{eq:ham:goni}
\end{equation}
where $\kappa$ parametrizes the  self-avoidance of the spin cluster boundaries.
The purely plaquette Hamiltonian with $\kappa=0$,
\begin{equation}
  \Ham = -\frac{1}{2}\plaqsum,
  \label{eq:ham:gonikappa0}
\end{equation}
allows spin cluster boundaries to intersect without energetic penalty. It has
attracted some attention recently, as it displays a strong first-order
transition~\cite{firstorder} and evidence of glass-like behaviour~\cite{glassy}
at low temperatures in spite of the absence of quenched disorder. 

The strong first-order nature of the transition for the purely plaquette
Hamiltonian has meant that its has proved difficult to obtain consistent values
for the inverse transition temperature. Only recently, it was understood that
the exponential degeneracy $q = 2^{3L}$~\cite{degeneracy} in the
low-temperature phase, for the periodic system living on a cube with linear lattice
size $L$, severely changes the finite-size corrections and leads to nonstandard
finite-size-scaling~\cite{mueller2014}.

First estimates {of} the inverse transition temperature were given by a
mean-field approximation that yielded $\beta^\infty = 0.325$ and early canonical
Monte Carlo simulations gave $\beta^\infty=0.50(1)$~\cite{Johnston1996}. {Later},
detailed finite-size scaling analysis of Monte Carlo data with {\it fixed}
boundary conditions found $\beta^\infty = 0.54757(63)$~\cite{Baig2004}.  More
recently, another value of $0.510(2)$, that is apparently compatible with the
first simulations~\cite{Johnston1996}, was suggested by analysing a dual
representation of the model with periodic boundary
conditions which turns out to be an anisotropic Ashkin-Teller
model~\cite{Johnston2011}. Here, two spins $\sigma, \tau$ live on each vertex
of a cubic lattice, with nearest-neighbour interactions along the $x,y$
and $z$-axes,
\begin{equation} 
  \Ham^d = -\frac{1}{2}\sum\limits_{\nn{i}{j}_x}\sigma_i\sigma_j -
  \frac{1}{2}\sum\limits_{\nn{i}{j}_y}\tau_i\tau_j
  -\frac{1}{2}\sum\limits_{\nn{i}{j}_z}\sigma_i\sigma_j\tau_i\tau_j.
  \label{eq:ham:dual}
\end{equation}

To resolve these discrepancies, we have conducted multicanonical
Monte Carlo simulations of the original model and its dual
representation. {In the remainder of the paper we present the results of these simulations
and the underlying theory of finite-size scaling at first order phase transitions used in extracting the conclusions.}

{
In section 2, after first discussing ``standard'' first-order finite-size scaling behaviour for models with
periodic boundary conditions, we  observe that  nonstandard first-order finite-size scaling behaviour will occur when there is an  exponentially large
degeneracy of the low-temperature phase, which is the case for both the plaquette Hamiltonian and
its dual. The scaling corrections for various observables are presented in detail in this case.}  

{
In section 3 we discuss the simulations themselves, for the plaquette Hamiltonian
with periodic boundary conditions, for the dual model with periodic boundary conditions and for the plaquette Hamiltonian
with fixed boundary conditions.   
We determine characteristic
quantities of these systems, such as the specific heat, Binder's energy
parameter, the latent heat and interface tension, as well as the inverse
temperature of the transition. We find for the first time, with the use of the nonstandard scaling relations, a consistent estimate
of the  inverse critical temperature from the plaquette Hamiltonian and its dual with periodic boundary conditions and the plaquette Hamiltonian with fixed boundary conditions. The self-consistency of the simulations is also confirmed by extracting various prefactors of scaling corrections both directly from fits and by calculation of energy moments.}

{
Finally, in section 4 we summarize our results for the measured physical quantities and note that the latent heat of the transition in the gonihedric model appears to be boundary condition dependent.}

\section{Finite-Size Scaling for First-Order Phase Transitions}

\label{sec:fss}
In spite of the ubiquity of first-order phase transitions~\cite{ubiquity} it
was only relatively recently that the initial studies of finite-size scaling
for first-order transitions were carried out~\cite{pioneer} and subsequently
pursued further in~\cite{furtherd}. Rigorous results for periodic boundary
conditions were derived in \cite{rigorous-fss,rigorous-fss-potts}.  For the
discussion of scaling laws under periodic boundary conditions here, we will
first employ the two-state model of~\cite{twostatemodel} which is capable of
correctly reproducing the prefactors of the leading contributions. To have this
paper reasonably self-contained, we will recall the principles and main results
in the following. In this model we assume the system {spends} some time in
either one of the $q$ ordered phases or in the disordered phase, where
transitions between the phases are instantaneous and fluctuations within the
phases are neglected. Let $W_{\rm o}$ denote the fraction of the total time {spent} in
one of the $q$ ordered phases and  $W_{\rm d} = 1 - W_{\rm o}$ {the fraction spent} in the
disordered phase. We associate energies $e_{\rm o}$ and $e_{\rm d}$ 
with the phases.
Under {these} assumptions, the energy moments are simply given by 
$\left<e^n\right> = W_{\rm o}e_{\rm o}^n + (1-W_{\rm o})e_{\rm d}^n$. 
The specific heat
$C_V(\beta, L) = -\beta^2\partial e(\beta, L)/\partial\beta$ as an expression
{in terms} of {these} moments becomes
\begin{equation}
  C_V(\beta, L) = L^d\beta^2\left(\left< e^2\right> - \left< e \right>^2\right) =
  L^d\beta^2 W_{\rm o}(1-W_{\rm o})\Delta e^2, 
  \label{eq:specheat}
\end{equation}
with $\Delta e = e_{\rm d} - e_{\rm o}$. 
Varying $W_{\rm o}$, we find the maximum
\begin{equation} 
  C^{\rm max} = L^d (\beta\Delta e/2)^2 \approx  L^d (\beta^\infty\Delta \hat{e}/2)^2 \label{eq:fss:specheatmax}
\end{equation} for $W_{\rm o} = W_{\rm d} = 0.5$, where the {ordered
and disordered} peaks of the energy probability density have equal weight. 
Here, we have assumed that $\beta$, $e_{\rm o}$ and $e_{\rm d}$ deviate from the
infinite-volume limit $\beta^\infty$, {$\hat{e}_{\rm o} = e_{\rm o}(\beta^\infty)$} and $\hat{e}_{\rm d}$,
respectively, only by terms of order {$1/L^d$}.

In close analogy, we can write the Binder parameter in terms of the two-state moments
\begin{eqnarray}
  B(\beta, L) &=& 1 - \frac{\langle e^4\rangle}{3\langle e^2 \rangle^2} = 1 - \frac{ W_{\rm o}\hat{e}_{\rm o}^4 + (1-W_{\rm o})\hat{e}_{\rm d}^4}{3\left(W_{\rm o}\hat{e}_{\rm o}^2 + (1-W_{\rm o})\hat{e}_{\rm d}^2\right)^2}\;,\label{eq:binder}
\end{eqnarray}
from which we can calculate the minimum with respect to the weights at $W_{\rm
o} = \hat{e}_{\rm d}^2/(\hat{e}_{\rm o}^2 + \hat{e}_{\rm d}^2)$, such that  
$W_{\rm o}/W_{\rm d} = \hat{e}_{\rm d}^2/\hat{e}_{\rm o}^2$ with a value of 
\begin{eqnarray}
  B^{\rm min}(L) = 1 - \frac{1}{12}\left( \frac{\hat{e}_{\rm o}}{\hat{e}_{\rm d}} +
  \frac{\hat{e}_{\rm d}}{\hat{e}_{\rm o}} \right)^2.\label{eq:fss:bindermin}
\end{eqnarray}

The free-energy densities, $f_{\rm o}$ and $f_{\rm d}$, of any one of the
ordered phases or the disordered phase govern their probability
\begin{equation}
  p_{\rm o}\propto {\rm e}^{-\beta L^d f_{\rm o}}\mbox{ and } p_{\rm d} \propto
  {\rm e}^{-\beta L^d f_{\rm d}}, 
\end{equation}
and the fraction of time spent in the ordered phases is proportional to $q p_{\rm o}$.
Neglecting {exponentially} small corrections in the linear lattice size \mbox{$L$~\cite{rigorous-fss,rigorous-fss-potts,twostatemodel}},
the ratio of the fraction of time spent in the respective phases is given by 
\begin{equation}
  W_{\rm o}/W_{\rm d} \simeq q{\rm e}^{- L^d \beta f_{\rm o}}/{\rm e}^{-\beta L^d f_{\rm d}} \; .
\end{equation}
Expanding the logarithm $\ln (W_{\rm o}/W_{\rm d}) = \ln q + L^d\beta(f_{\rm d} - f_{\rm o})$ around the infinite-volume phase-transition temperature $\beta^\infty$
 leads to
\begin{equation}
  \ln (W_{\rm o}/W_{\rm d}) = \ln q + L^d\Delta \hat{e}(\beta - \beta^\infty) + \dots \, ,
\end{equation}
which after truncation {can be solved} for the inverse temperatures $\beta$. 
For $W_{\rm o} = W_{\rm d} = 0.5$ we find the inverse temperature $\beta^{\rm eqw}$, where both peaks of the energy probability density have equal weight, and, to leading order, the location $\beta^{C_V^{\rm max}}$ of the specific heat maximum,
\begin{equation} 
 \beta^{C_V^{\rm max}}(L) = \beta^{\rm eqw}(L) = \beta^\infty - \frac{\ln q}{\Delta\hat{e}L^{d}} + \dots \, .
\end{equation}
The minimum of Binder's parameter at $W_{\rm o}/W_{\rm d} = \hat{e}_{\rm
d}^2/\hat{e}_{\rm o}^2$ is located at the inverse temperature
\begin{eqnarray}
  \beta^{B^{\rm min}}(L) = \beta^\infty - \frac{\ln(q\hat{e}^2_{\rm
  o}/\hat{e}^2_{\rm d})}{L^d\Delta \hat{e}}  + \dots \, .
\end{eqnarray}

In spite of its simplicity the model captures the
essential features of first-order phase transitions and, importantly for our purposes, correctly predicts the
prefactors of the leading finite-size scaling corrections for a class of models
with a contour representation, such as the $q$-state Potts model, where a
completely rigorous theory of scaling also exists~\cite{rigorous-fss-potts}.
This {rigorous theory} enables the calculation of the coefficients of higher-order terms in a 
systematic asymptotic expansion in powers of $1/L^d$~\cite{twostatemodel,lee-kosterlitz}.
In addition, there are further corrections that decay exponentially fast with 
growing system size~\cite{exponentialcorr}.

These models for periodic boundary conditions have, up to {exponentially small}
corrections in $L$, a canonical partition function of the form{~\cite{rigorous-fss-potts}}
\begin{equation}
  Z\left(\beta, L\right) = q {\rm e}^{-\beta L^d f_o(\beta)} + {\rm e}^{-\beta L^d f_d(\beta)},
\end{equation} 
allowing a more rigorous derivation of inverse transition temperatures.
The inverse temperature of equal peak weight then reads{~\cite{twostatemodel}}
\begin{eqnarray} 
  \beta^{\rm eqw}(L) 
    &\!\!=&\!\! \beta^\infty\! - \beta^\infty\frac{\ln q}{\Delta\hat{s}L^{d}} 
        + \beta^\infty\left( \frac{\ln q}{\Delta\hat{s}L^d}\right)^2
        \left(\frac{\Delta\hat{C}}{2\Delta\hat{s}} \right)
        + {\cal O}\left(\frac{(\ln q)^3}{L^{3d}} \right)\!,\label{eq:fss:beta:eqw}
\end{eqnarray}
where $\Delta\hat{s} = \beta^{\infty}\Delta\hat{e}$ is the transition entropy
and $\Delta\hat{C} = \hat{C}_d - \hat{C}_o$.  For the location of the
specific-heat maximum and the minimum of the Binder parameter one finds
{~\cite{twostatemodel,lee-kosterlitz}}
\begin{eqnarray} 
	\beta^{C^{\rm max}_V}(L) 
    &=& \beta^\infty - \beta^\infty\frac{\ln q}{\Delta\hat{s}L^{d}} 
        + \beta^\infty\left( \frac{1}{\Delta\hat{s}L^d}\right)^2 
        \left(\frac{\Delta\hat{C}}{2\Delta\hat{s}}\left( \left(\ln q\right)^2 -12  \right) 
      + 4\right) \nonumber\\
      && +\ {\cal O}\left(\frac{(\ln q)^3}{L^{3d}} \right), \label{eq:fss:beta:specheat}\\
  \beta^{B^{\rm min}}(L) 
    &=& \beta^\infty - \beta^\infty\frac{\ln(q\hat{e}^2_{\rm
  o}/\hat{e}^2_{\rm d})}{L^d\Delta \hat{s}}  + \frac{a_2}{L^{2d}} + 
  {\cal O}\left( \frac{(\ln(q\hat{e}^2_{\rm o}/\hat{e}^2_{\rm d}))^3}{L^{3d}}\right),
  \label{eq:fss:beta:binder}
%
\end{eqnarray}
where the {expression} for $a_2$ is {explicitly} known but very
complicated~\cite{twostatemodel}, and will simplify in our special case (see
below).

The leading correction {to} the inverse temperature of equal peak height,
$\beta^{\rm eqh}$, comes {more} heuristically from a double gaussian approximation of
the energy probability density~\cite{twostatemodel},
\begin{equation} 
  \beta^{\rm eqh}(L) = \beta^\infty - \beta^\infty\frac{\ln (q \hat{C}_{\rm d}/\hat{C}_{\rm o})}{2\Delta\hat{s}L^{3}} + {\cal O}\left(\frac{(\ln (q \hat{C}_{\rm d}/\hat{C}_{\rm o}))^2}{L^{6}}\right).
\end{equation} 

Usually, the low-temperature degeneracy $q$ is a constant
{(like in a $q$-state Potts model)} and standard
finite-size scaling behaviour at a first-order transition displays a leading
contribution proportional to the inverse volume $L^{-d}$.  However, for the
three-dimensional gonihedric Ising model~\eref{eq:ham:goni}, the degeneracy $q$
is {exponentially} dependent on the linear system size. 
By construction, the model shows a highly degenerate ground-state for
all parameters $\kappa$. In the special case of vanishing energetic penalty for
self-intersecting spin cluster boundaries, $\kappa=0$, the degeneracy, 
\begin{equation}
	q = 2^{3L} = e^{3 L \ln 2}, 
\label{eq:degeneracy}
\end{equation}
is apparent even for finite temperatures~\cite{degeneracy}.
{
{The} usual $1/L^{3}$ behaviour {is} {therefore} transmuted to {a} $1/L^{2}$ behaviour~\cite{mueller2014}. {Furthermore}, the 
{finite-size scaling} corrections {to $\beta^{\rm eqw}(L)$} in \eref{eq:fss:beta:eqw} and in the scaling {law} 
\eref{eq:fss:beta:specheat} {for $\beta^{C^{\rm max}_V}(L)$} now coincide up to  order ${\cal O}(L^{-4})$,}
\begin{eqnarray} 
  \beta^{C^{\rm max}_V}(L) \approx \beta^{\rm eqw}(L) 
    &=& \beta^\infty - \frac{\ln 2^{3L}}{\Delta\hat{e}L^{3}}
        + \frac{\Delta\hat{C}}{2\Delta\hat{e}}
          \left( \frac{\ln 2^{3L}}{\Delta\hat{s}L^3}\right)^2
        + {\cal O}\left(\frac{(\ln q)^3}{L^9} \right)\nonumber \\
    &=& \beta^\infty - \frac{3\ln 2}{\Delta\hat{e}L^{2}}
        + \frac{\Delta\hat{C}}{2\Delta\hat{e}} 
          \left( \frac{3\ln 2}{\Delta\hat{s}L^2}\right)^2
        + {\cal O}(L^{-6})\label{eq:fss:beta:eqw2}\label{eq:fss:beta:specheat2}\,.
\end{eqnarray}
The scaling {law} for the peak location of Binder's 
parameter~\eref{eq:fss:beta:binder} becomes
\begin{eqnarray} 
  \beta^{B^{\rm min}}(L) 
    &\!=&\! \beta^\infty - \frac{\ln(2^{3L}\hat{e}^2_{\rm o}/
        \hat{e}^2_{\rm d})}{\Delta \hat{e}L^3}  + 
        \frac{\Delta\hat{C}}{2\Delta\hat{e}} 
        \left( \frac{\ln 2^{3L}}{\Delta\hat{s}L^3}\right)^2 + \dots \nonumber\\ 
    &\!=&\! \beta^\infty - \frac{3\ln2}{\Delta \hat{e}L^2}
        - \frac{\ln(\hat{e}^2_{\rm o}/\hat{e}^2_{\rm d})}{\Delta\hat{e}L^3} 
        + \frac{\Delta\hat{C}}{2\Delta\hat{e}} 
          \left( \frac{3\ln 2}{\Delta\hat{s}L^2}\right)^2\!\! + {\cal O}(L^{-6}),
  \label{eq:fss:beta:binder2} 
\end{eqnarray}
where we have used the fact that only the contribution to $a_2$ with the highest
power of $\ln q$, $a_2 = (\ln q/\Delta\hat{s})^2 \Delta\hat{C}/2\Delta\hat{e} + \dots$, 
{contributes} to the order given.  
The leading contribution {to} the finite-size correction is thus also proportional to 
$L^{-2}$, and the prefactor of the contribution ${\cal O}(L^{-4})$ becomes the 
same {as that} found for the inverse temperatures of {the} equal peak weight and the peak 
location of the specific heat. {Note that there is, however, an additional correction term of ${\cal O}(L^{-3})$.}

The leading correction {to} the inverse temperature of equal peak height,
$\beta^{\rm eqh}$, is now also {of} order ${\cal O}\left( L^{-2} \right)$,
\begin{equation} 
  \beta^{\rm eqh}(L) = \beta^\infty - \frac{3\ln 2}{\Delta\hat{e}L^{2}} - \frac{\ln(\hat{C}_{\rm d}/\hat{C}_{\rm o})}{2\Delta\hat{e}L^{3}} + {\cal O}(L^{-4}).
  \label{eq:fss:beta:eqh} 
\end{equation} 

The extremal values of the specific heat and Binder's parameter change with
the system size according to
\begin{eqnarray} 
  C^{\rm max}(L) 
    &\!\!=&\!\!\left( \frac{\Delta\hat{s}}{2}\right)^2\!L^3 +
        \frac{\ln q(\Delta{\hat{C}} - \Delta{\hat{s}})}{2} + 
        \frac{\hat{C}_{\rm d} + \hat{C}_{\rm o}}{2} + \dots\nonumber\\
    &\!\!=&\!\! \left( \frac{\Delta\hat{s}}{2}\right)^2\!L^3 +
        \frac{3\ln 2(\Delta{\hat{C}}-\Delta{\hat{s}})}{2}L + 
        \frac{\hat{C}_{\rm d} + \hat{C}_{\rm o}}{2} + {\cal O}(L^{-1}) \label{eq:fss:specheat}
\end{eqnarray} 
and 
\begin{equation} 
  B^{\rm min}(L) = 1 - \frac{1}{12}\left( \frac{\hat{e_{\rm o}}}{\hat{e_{\rm d}}} +
  \frac{\hat{e_{\rm d}}}{\hat{e_{\rm o}}} \right)^2  + aL^{-2} + {\cal O}(L^{-3}).
  \label{eq:fss:binder} 
\end{equation}
The {prefactor in the} first correction {for $B^{\rm min}(L)$} reads
\begin{equation}
  a = \left( \frac{1}{\beta^\infty} \right)^2  \frac{ 3\ln 2(\hat{e}_{\rm d} + \hat{e}_{\rm o}) (\hat{C}_{\rm o} \hat{e}_{\rm d} - \hat{C}_{\rm d} \hat{e}_{\rm o}) (\hat{e}_{\rm d}^2 + \hat{e}_{\rm o}^2)}{6\hat{e}_{\rm d}^3 \hat{e}_{\rm o}^3},\label{eq:fss:binder:a2}
\end{equation}
which comes from an even more complicated expression {of} ${\cal O}(L^{-3})$ 
in the general case~\cite{twostatemodel}. {Here} we {have} already {taken} the degeneracy $q=2^{3L}$ into account.

The Taylor series of the energy in the {ordered and disordered} phases, $e_{\rm o/d}$, around $\beta {=}\beta^\infty$ reads
\begin{equation} 
  e_{\rm o/d}(\beta) = \hat{e}_{\rm o/d} + \frac{\partial e_{\rm o/d}}{\partial \beta}\bigg|_{\beta=\beta^\infty}(\beta - \beta^\infty) + {\cal O}\left(\left( \beta - \beta^\infty \right)^2 \right), 
\end{equation}
where the specific heat of the {ordered and disordered} phase enters the leading correction. Calculating the energies at inverse temperature $\beta^{\rm eqw}$, the scaling of the energy {fulfils}
\begin{equation} 
  e_{\rm o/d}(\beta^{\rm eqw}) = \hat{e}_{\rm o/d} + \hat{C}_{\rm o/d}\left(\frac{1}{\beta^\infty}\right)^2(\beta^{\rm eqw} - \beta^\infty) + {\cal O}\left(\left( \beta^{\rm eqw} - \beta^\infty \right)^2\right).
\end{equation}
The difference $\beta^{\rm eqw} - \beta^\infty$ is known from \eref{eq:fss:beta:eqw2}, therefore the expression
\begin{equation} 
  e_{\rm o/d}(\beta^{\rm eqw}) = \hat{e}_{\rm o/d} + \left(\frac{1}{\beta^\infty}\right)^2\hat{C}_{\rm o/d} \frac{3\ln(2)}{\Delta\hat{e} L^2} + {\cal O}\left( L^{-4} \right)
  \label{eq:fss:energies}
\end{equation}
shows the {finite-size} corrections {to} the energy.

The same change of leading contributions is
apparent in the dual model~\eref{eq:ham:dual}, where a similar low-temperature
phase degeneracy is expected (but, in contrast to the original model, not
proven). These considerations will also apply to other models with periodic
boundary conditions which have a degeneracy that depends {exponentially} on the system size. 

For fixed boundary conditions, surface effects play an important
role~\cite{fssfbc}. Here, the inverse transition temperature of the gonihedric
Ising model is shifted by a leading term of order $\mathcal{O}(L^{-1})$ for
finite lattices of linear lattice size $L$. Thus in this case we expect
\begin{equation} 
\beta(L) =\beta^\infty  - \frac{a_1}{L} + {\mathcal O}(L^{-2}) 
 \end{equation}
for the peak locations of both the specific heat and Binder's parameter. 

\section{Simulation Results} 
An effective way of combating supercritical slowing down near first-order phase
transitions, where canonical simulations tend to get trapped in either the
disordered or ordered phases, is to use the multicanonical Monte Carlo
algorithm~\cite{muca,mucajanke}. At such first order transitions cooling a system down or
heating it up also makes estimation of the transition temperature inherently
difficult using standard algorithms due to hysteresis effects. In a
multicanonical simulation it is possible to systematically improve guesses of
the energy probabilty distribution before the actual production run by using
recursive estimates~\cite{mucaweights}. 

The rare states lying between the disordered and ordered phases in the energy
histogram are then promoted artificially, decreasing the autocorrelation time
and allowing the system to oscillate more rapidly between phases. Canonical
estimators can then be retrieved by weighting the multicanonical data to yield
Boltzmann-distributed energies. Such reweighting techniques are very powerful
when combined with the multicanonical simulations, allowing the calculation of
observables over a broad range of temperatures.

\subsection{Observables}

Standard observables such as the specific heat
\eref{eq:specheat} 
and Binder's energy parameter~\eref{eq:binder} were calculated at
different temperatures from our data for both the gonihedric Ising
model~\eref{eq:ham:gonikappa0} and its dual~\eref{eq:ham:dual}. The
positions of their peaks, $\beta^{C_V^{\rm max}}(L)$ and $\beta^{B^{\rm
min}}(L)$ were then determined by the reweighting techniques~\cite{reweighting}.

Other estimates of the inverse critical temperature are given by $\beta^{\rm eqw}$ and
$\beta^{\rm eqh}$, where the disordered and ordered peaks of the energy
probability density $p(e)$ have the same weight or height,  respectively. In
practice, we use reweighting techniques to get an estimator of the energy
probability densities at different temperatures. $\beta^{\rm eqw}$ is  chosen
systematically to minimize 
\begin{equation} 
  D^{\rm eqw}(\beta) = \left( \sum_{e < e_{\rm min}} p(e, \beta) - \sum_{e \geq
  e_{\rm min}} p(e, \beta) \right)^2 
\end{equation}
where the energy of the minimum between both peaks, $e_{\rm min}$, is
determined beforehand to distinguish between the different phases.
The location of the minimum, $\beta^{\rm eqw}$, is then used to calculate the
energy moments of the {ordered and disordered} phases, 
\begin{eqnarray} 
  e^k_{\rm o}(L) = \sum_{e < e_{\rm min}} e^k\, p(e, \beta^{\rm eqw}) \Big/ \sum_{e < e_{\rm min}} p(e, \beta^{\rm eqw}),\nonumber\\
  e^k_{\rm d}(L) =
  \sum_{e \geq e_{\rm min}} e^k\, p(e, \beta^{\rm eqw}) \Big/\sum_{e \geq e_{\rm min}} p(e, \beta^{\rm eqw}),
  \label{eq:e-eqw}
\end{eqnarray} 
where $e_{\rm o/d}(L) = e^1_{\rm o/d}(L)$ is the energy in the respective phases, and their difference is an estimator of the latent heat $\Delta e(L)=e_{\rm d}(L)-e_{\rm o}(L)$. Also, the second and first moment combine to give the specific heat of the disordered and ordered phases, 
\begin{equation} 
  C_{\rm o/d}(L) = \beta^2L^d\left( e^2_{\rm o/d}(L) - \left(e^1_{\rm o/d}(L)\right)^2\right).
\end{equation}

To find the inverse transition temperature where both phases have equal
height we minimize 
\begin{equation} 
  D^{\rm eqh}(\beta) =  \left( \max_{e < e_{\rm min}}\{p(e, \beta)\} - \max_{e
  \geq e_{\rm min}}\{p(e, \beta)\} \right)^2, 
\end{equation} 
{as a function of $\beta$.}
The probability density $p(e, \beta^{\rm eqh})$ itself
at $\beta^{\rm eqh}$ is also of interest since one can make use of it to extract the reduced interface tension
\begin{equation} 
  \sigma(L) = \frac{1}{2L^2} \ln \left( \frac { \max \{ p(e, \beta^{\rm eqh})
  \} } { \min \{ p(e, \beta^{\rm eqh}) \} } \right),  
  \label{eq:interface-tension}
\end{equation} 
for periodic boundary conditions. This characteristic quantity of first-order
phase transitions is almost impossible to extract reliably from canonical
Metropolis simulations, since getting reasonable statistics on the suppressed
states is very hard. Multicanonical simulations, on the other hand, are
perfectly tailored for measurements of such rare events.

\subsection{Original plaquette model with periodic boundary conditions} 

\begin{figure}[hbt] 
  \begin{center} 
    \includegraphics{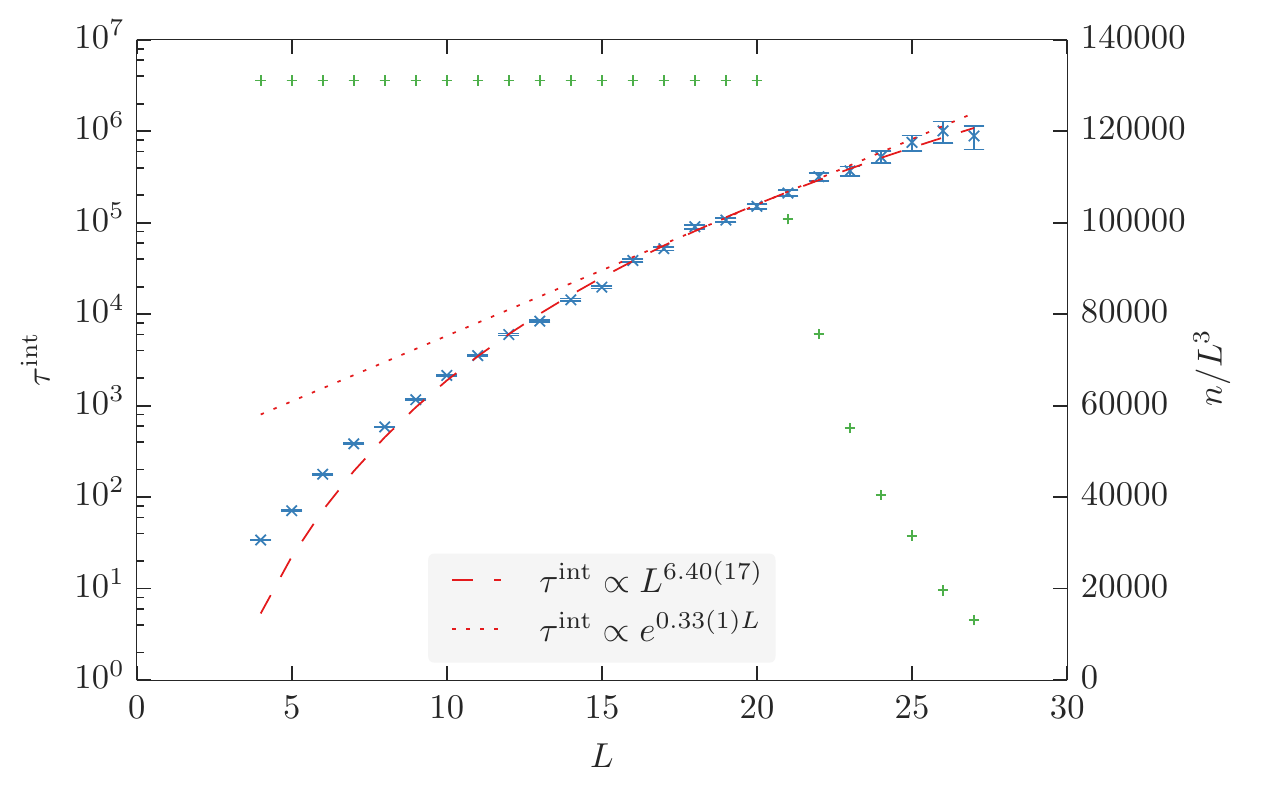} 
    \caption{
      The integrated autocorrelation time $\tau^{\rm int}$ in units of sweeps for
      the original model with periodic boundary conditions (blue markers; left
      axis) over the linear lattice size $L$. The dashed and dotted curves are 
      fits {with} a power law or an exponential law for lattices sizes $L\geq 16$.
      On the right axis, with green markers, the total number
      of {sweeps} $n$ {divided by the volume} is shown.}
    \label{fig:tauorig} 
  \end{center} 
\end{figure}

The quality of the simulations for the original plaquette
model~\eref{eq:ham:gonikappa0} with periodic boundary conditions {can be} judged 
by the integrated autocorrelation time $\tau^{\rm int}$ and the number of
{sweeps}
in Figure~\ref{fig:tauorig}. Here, $\tau^{\rm int}$ has been
determined with a self-consistent cutoff at $6\tau^{\rm int}$ and the error
comes from the known formula for this algorithm~\cite{dataanalysis},
$\epsilon_{\tau^{\rm int}} = \left( \tau^{\rm int} \right)^{3/2}\left(
\frac{24}{n} \right)^{1/2}$, where {$n$ is the number of measurements ($=$ number of sweeps when performing measurements every sweep).}
We would expect that the integrated autocorrelation time with perfect
multicanonical weights should in principle
increase linearly with the volume,
$\tau^{\rm int} \propto L^3$.

Error-weighted nonlinear least-squares fits 
{of a power law, $\tau^{\rm int} \propto L^{\alpha}$, 
to the actual measured
integrated autocorrelation times}
yield
{much larger} exponents $\alpha \approx 6.40(17)$ that vary a bit around $6.0$ depending on
the lattice sizes included in the fits.  Also for those fits with acceptable
$\chi^2 \approx 1$ that only include lattices with size $L\geq 16$, fits to an
exponential growth with $L$ are of comparable quality, {see Figure~\ref{fig:tauorig}}. With least-squares {fits} and no proper model
testing, we cannot distinguish between the two alternatives~\cite{clauset2009}.
 {In any case}, we find that the autocorrelation time grows significantly faster than
expected, an effect that may be attributed to free-energy barriers. 
Such hidden barriers appear for instance in droplet
condensation~\cite{freeEbarriers}, whose analog {with} the gonihedric Ising
model is, however, still unclear.

\begin{figure}[htb] 
  \begin{center} 
    \includegraphics[scale=0.98]{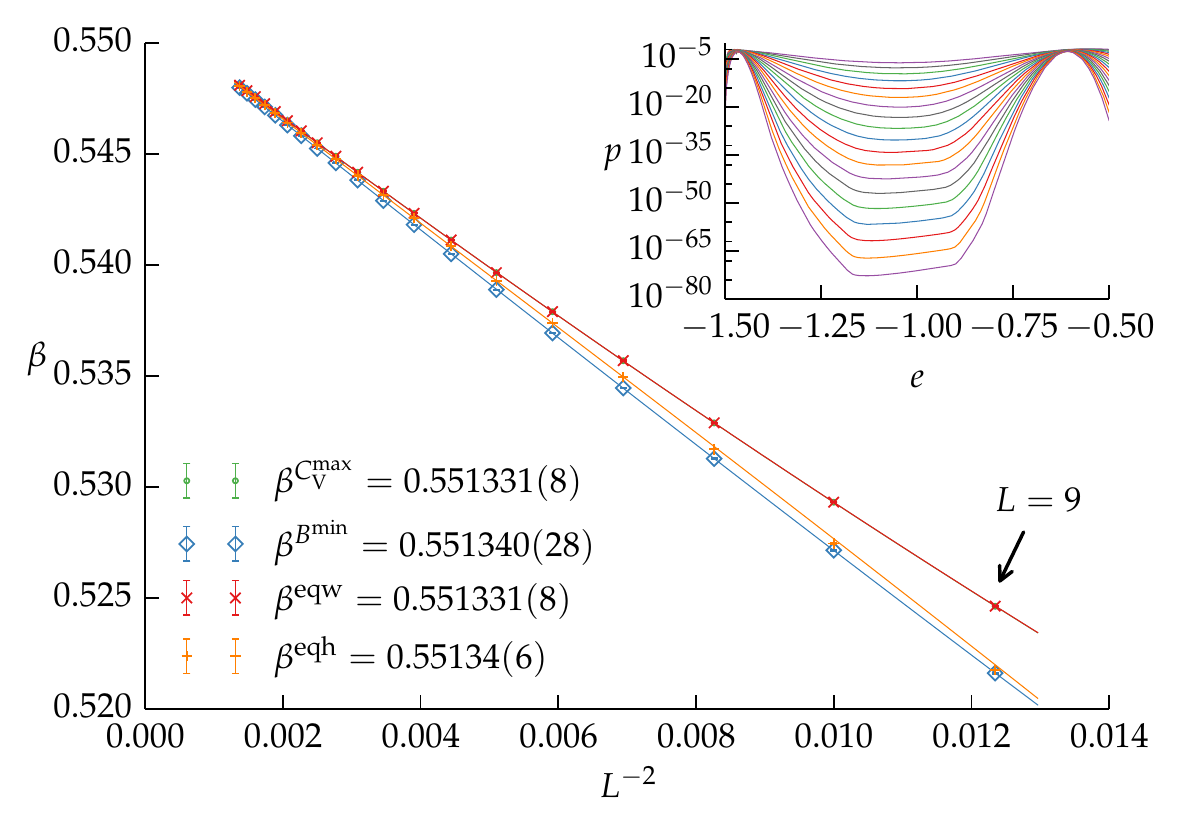} 
    \caption{
      Best fits up to order ${\cal O}(L^{-4})$ obtained for the original model
      with periodic boundary conditions (cf. Table~\ref{tab:resultingfits})
      using the (finite lattice) peak
      locations for the specific heat $C_V^{\rm max}$, Binder's energy
      parameter $B^{\rm min}$; or inverse temperatures $\beta^{\rm eqw}$ and
      $\beta^{\rm eqh}$, where the two peaks of the energy probability density
      are of same weight or have equal height, respectively. The values for
      $\beta^{\rm eqw}$ and $\beta^{C^{\rm max}_V}$ are indistinguishable in
      the plot. The inset shows the energy probability density $p(e)$ over $e =
      E/L^d$ at $\beta^{\rm eqh}$ for lattices with linear length $L\in \left\{
      9, 10, \dots, 27 \right\}$. }
    \label{fig:fit:orig-pbc} 
  \end{center} 
\end{figure}

For each lattice, we performed a maximum number of 
$n^{\rm max}=2^{17}\,L^3=131072\,L^3$ sweeps with an upper bound on the
computer time of around $500$ hours of real time for each lattice size. All
lattices with size $L\leq 20$ finished the desired number of sweeps, the larger
lattices were aborted after $500$ hours and collected {correspondingly} less statistics.  The
ratio $n/\tau^{\rm int}$ is a {measure} for the number of effectively
uncorrelated {data}. Although the autocorrelation time increases 
dramatically with the system size, the simulation of the largest lattice of
$V=27^3$ spins still effectively flipped more than 250 times between the two 
phases during the simulation. This is remarkable, given that rare states were 
suppressed by more than 60 orders of magnitude compared to the most probable 
states (see the inset of Figure~\ref{fig:fit:orig-pbc}). 

From our multicanonical data, we have extracted the resulting quantities of 
interest for different lattice sizes and listed them in 
Table~\ref{tab:originalresults}, where errors have been extracted
by jackknife analysis using $20$ blocks for each lattice size~\cite{jackknife}.  

We find that the inverse temperatures of the specific-heat maximum
$\beta^{C_V^{\rm max}}$ and of equal peak weights $\beta^{\rm eqw}$ {fall nearly}
together for all lattice sizes. This {is accounted for by the fact} that the 
higher-order corrections of order ${\cal O}\left(L^{-4}\right)$ in the scaling
{law \eref{eq:fss:beta:eqw2}} for these quantities
collapse {due to} the {exponential} degeneracy of the low-temperature phase
to induce the same prefactor.

\begin{landscape} \begin{table}[htpb] \tablesize \centering 
      \caption{Resulting quantities for the gonihedric Ising
        model~\eref{eq:ham:gonikappa0} with \emph{periodic\/} boundary conditions:
      extremal values for the specific heat $C_V^{\rm max}$, Binder's energy
      parameter $B^{\rm min}$, with their respective pseudo-critical inverse
      temperatures $\beta$, and temperatures where peaks of the energy
      probability density have equal heights and weights for finite lattices
      with linear length $L$. The finite-lattice interface tension is listed as
      $\sigma$, the energy of the ordered and disordered phases and their
      difference as $e_{\rm o}, e_{\rm d}$ and $\Delta e$. 
      The infinite lattice limits are listed as parameters of fits whose
      goodness-of-fit value $Q$ is given. Highlighted in light grey are the 
      datapoints used for fits with only leading order correction (lo).
      Additional datapoints used for fitting with subleading corrections (so)
      up to and including order ${\cal O}(L^{-4})$ are marked in dark grey.}
    \vspace{3ex} \begin{tabular}{l*{10}{r}} \toprule
    \multicolumn{1}{c}{$L$} & \multicolumn{1}{c}{$\beta^{C_V^{\rm max}}$} &
    \multicolumn{1}{c}{$C_V^{\rm max}$} & \multicolumn{1}{c}{$\beta^{B^{\rm
    min}}$} & \multicolumn{1}{c}{$B^{\rm min}$} &
    \multicolumn{1}{c}{$\beta^{\rm eqw}$} & \multicolumn{1}{c}{$\beta^{\rm
    eqh}$} & \multicolumn{1}{c}{$\sigma$} & \multicolumn{1}{c}{$e_{\rm o}$} &
    \multicolumn{1}{c}{$e_{\rm d}$} & \multicolumn{1}{c}{$\Delta e$} \\
    \midrule
08&            $0.518228(26)$ &               $27.061(7)$ &            $0.513850(27)$ &             $0.25211(16)$ &            $0.518244(26)$ &            $0.514007(23)$ &              $0.05659(6)$ &            $-1.43921(13)$ &            $-0.56223(14)$ &             $0.87698(11)$ \\
09&      \hlg{$0.524636(24)$} &               $39.611(6)$ &      \hlg{$0.521626(24)$} &             $0.26024(14)$ &      \hlg{$0.524644(24)$} &            $0.521769(24)$ &              $0.06240(5)$ &             $-1.44791(5)$ &             $-0.56529(8)$ &              $0.88262(8)$ \\
10&      \hlg{$0.529322(18)$} &               $55.068(7)$ &      \hlg{$0.527159(19)$} &             $0.27023(12)$ &      \hlg{$0.529327(18)$} &            $0.527463(18)$ &            $0.067756(29)$ &           $-1.452764(28)$ &             $-0.57011(7)$ &              $0.88265(7)$ \\
11&      \hlg{$0.532894(13)$} &               $73.766(7)$ &      \hlg{$0.531286(13)$} &              $0.27998(7)$ &      \hlg{$0.532897(13)$} &            $0.531705(12)$ &              $0.07305(4)$ &     \hlg{$-1.455896(28)$} &             $-0.57525(5)$ &              $0.88065(4)$ \\
12&      \hlg{$0.535696(13)$} &               $96.094(8)$ &      \hlg{$0.534467(13)$} &              $0.28847(8)$ &      \hlg{$0.535698(13)$} &     \hllg{$0.534965(13)$} &              $0.07804(4)$ &     \hlg{$-1.458205(19)$} &             $-0.57996(5)$ &              $0.87824(5)$ \\
13&      \hlg{$0.537902(12)$} &             $122.315(10)$ &     \hllg{$0.536941(12)$} &              $0.29593(8)$ &      \hlg{$0.537903(12)$} &     \hllg{$0.537383(12)$} &              $0.08285(4)$ &     \hlg{$-1.459850(21)$} &             $-0.58424(5)$ &              $0.87561(5)$ \\
14&      \hlg{$0.539662(10)$} &             $152.801(12)$ &     \hllg{$0.538897(10)$} &              $0.30221(6)$ &      \hlg{$0.539663(10)$} &      \hllg{$0.539297(9)$} &            $0.087156(24)$ &     \hlg{$-1.461138(15)$} &       \hlg{$-0.58795(4)$} &              $0.87319(4)$ \\
15&       \hlg{$0.541128(9)$} &             $187.897(12)$ &      \hllg{$0.540508(9)$} &              $0.30756(4)$ &       \hlg{$0.541128(9)$} &      \hllg{$0.540853(9)$} &              $0.09105(4)$ &     \hlg{$-1.462154(20)$} &     \hlg{$-0.591175(23)$} &            $0.870979(30)$ \\
16&      \hlg{$0.542329(10)$} &             $227.917(10)$ &     \hllg{$0.541820(10)$} &              $0.31207(5)$ &      \hlg{$0.542329(10)$} &      \hllg{$0.542114(9)$} &              $0.09430(4)$ &      \hlg{$-1.462961(8)$} &     \hlg{$-0.593953(30)$} &            $0.869009(29)$ \\
17&      \hllg{$0.543326(8)$} &             $273.174(11)$ &      \hllg{$0.542903(8)$} &              $0.31597(5)$ &      \hllg{$0.543326(8)$} &      \hllg{$0.543151(8)$} &            $0.096981(22)$ &      \hlg{$-1.463630(8)$} &     \hlg{$-0.596392(28)$} &            $0.867238(24)$ \\
18&      \hllg{$0.544181(9)$} &              $324.070(9)$ &      \hllg{$0.543825(9)$} &              $0.31923(4)$ &      \hllg{$0.544181(9)$} &      \hllg{$0.544035(9)$} &              $0.09928(4)$ &     \hllg{$-1.464179(9)$} &     \hlg{$-0.598459(27)$} &            $0.865720(21)$ \\
19&     \hllg{$0.544904(10)$} &             $380.852(10)$ &     \hllg{$0.544602(10)$} &              $0.32210(4)$ &     \hllg{$0.544904(10)$} &     \hllg{$0.544781(10)$} &              $0.10128(4)$ &     \hllg{$-1.464630(9)$} &     \hlg{$-0.600290(25)$} &            $0.864339(21)$ \\
20&      \hllg{$0.545510(5)$} &             $443.910(12)$ &      \hllg{$0.545252(5)$} &     \hllg{$0.324501(26)$} &      \hllg{$0.545511(5)$} &      \hllg{$0.545403(5)$} &            $0.102911(25)$ &     \hllg{$-1.465009(6)$} &     \hlg{$-0.601844(17)$} &            $0.863165(16)$ \\
21&      \hllg{$0.546044(7)$} &      \hllg{$513.571(12)$} &      \hllg{$0.545821(7)$} &     \hllg{$0.326601(29)$} &      \hllg{$0.546044(7)$} &      \hllg{$0.545952(7)$} &            $0.104440(20)$ &     \hllg{$-1.465339(9)$} &    \hllg{$-0.603216(21)$} &     \hllg{$0.862124(15)$} \\
22&      \hllg{$0.546500(6)$} &      \hllg{$590.141(19)$} &      \hllg{$0.546306(6)$} &     \hllg{$0.328422(24)$} &      \hllg{$0.546500(6)$} &      \hllg{$0.546420(6)$} &       \hllg{$0.10576(4)$} &     \hllg{$-1.465615(6)$} &    \hllg{$-0.604412(15)$} &     \hllg{$0.861203(16)$} \\
23&      \hllg{$0.546914(8)$} &      \hllg{$673.971(21)$} &      \hllg{$0.546745(8)$} &       \hllg{$0.33005(4)$} &      \hllg{$0.546914(8)$} &      \hllg{$0.546843(8)$} &       \hllg{$0.10702(5)$} &     \hllg{$-1.465856(7)$} &    \hllg{$-0.605491(23)$} &     \hllg{$0.860365(20)$} \\
24&      \hllg{$0.547270(9)$} &      \hllg{$765.339(21)$} &      \hllg{$0.547121(9)$} &       \hllg{$0.33152(4)$} &      \hllg{$0.547270(9)$} &      \hllg{$0.547207(9)$} &       \hllg{$0.10819(7)$} &     \hllg{$-1.466076(9)$} &    \hllg{$-0.606468(24)$} &     \hllg{$0.859607(21)$} \\
25&      \hllg{$0.547584(9)$} &      \hllg{$864.753(27)$} &      \hllg{$0.547452(9)$} &       \hllg{$0.33271(4)$} &      \hllg{$0.547584(9)$} &      \hllg{$0.547528(9)$} &       \hllg{$0.10901(5)$} &     \hllg{$-1.466270(6)$} &    \hllg{$-0.607270(23)$} &     \hllg{$0.858999(20)$} \\
26&     \hllg{$0.547856(14)$} &        \hllg{$972.36(5)$} &     \hllg{$0.547739(14)$} &       \hllg{$0.33376(6)$} &     \hllg{$0.547856(14)$} &     \hllg{$0.547805(14)$} &       \hllg{$0.10997(9)$} &    \hllg{$-1.466413(11)$} &      \hllg{$-0.60798(4)$} &       \hllg{$0.85844(4)$} \\
27&     \hllg{$0.548099(14)$} &       \hllg{$1088.54(4)$} &     \hllg{$0.547994(14)$} &       \hllg{$0.33475(6)$} &     \hllg{$0.548099(14)$} &     \hllg{$0.548053(14)$} &      \hllg{$0.11066(10)$} &    \hllg{$-1.466578(13)$} &      \hllg{$-0.60865(4)$} &     \hllg{$0.857927(28)$} \\
\midrule
\rowcolor{lightgrey} lo &           $0.551233(10)$&         $0.055072(4)L^3$&            $0.551350(6)$&             $0.34729(7)$&           $0.551233(10)$&            $0.551277(5)$&            $0.12037(18)$&          $-1.468500(11)$&            $-0.61701(7)$&             $0.85148(5)$\\
\rowcolor{lightgrey}  Q &                     0.54&                     0.35&                     0.72&                     0.50&                     0.54&                     0.90&                     0.38&                     0.56&                     0.56&                     0.59\\
\rowcolor{grey}      so &            $0.551331(8)$&                         &           $0.551340(27)$&                         &            $0.551331(8)$&             $0.55134(6)$&                         &          $-1.468373(12)$&            $-0.61771(6)$&                         \\
\rowcolor{grey}       Q &                     0.95&                         &                     0.92&                         &                     0.95&                     0.93&                         &                     0.63&                     0.59&                         \\

\bottomrule \end{tabular}
\label{tab:originalresults} \end{table} \end{landscape}

{Least-squares fits of} the data in Table~\ref{tab:originalresults}
according to the laws in section~\ref{sec:fss} have been conducted.  
We have left out the smaller lattices systematically for each fit,
until a goodness-of-fit value of at least $Q = 0.5$ was found 
for each observable individually.  
The same protocol was employed earlier~\cite{mueller2014} for a reduced time
series, where only every $L^3$-th measurement was used. There we
were not challenging the prefactors of higher-order corrections {so}  
the reduced time series was sufficient.
We list all the fit parameters obtained for {both} the full time series and the
reduced one in Table~\ref{tab:resultingfits} along with the quality-of-fit
parameters $Q$ and the degrees of freedom left.
The inverse transition temperatures in the thermodynamic limit
are {effectively identical} and do not depend on whether we use the reduced or the
full dataset {or} on the precise final averaging procedure. A graphical
representation of the best fits is given in
Figure~\ref{fig:fit:orig-pbc}.

\begin{table}[btp] 
  \tablesize 
  \centering 
  \caption{Resulting parameters of the best fits to the extremal points $\beta$
    for the specific heat $C_V^{\rm max}$, Binder's energy parameter $B^{\rm
    min}$; or inverse temperatures $\beta^{\rm eqw}$ and $\beta^{\rm eqh}$ to
    laws of the form $\beta(L) = \beta^\infty + p_2/L^2 + p_3/L^3 + p_4/L^4$.
    Parameters $p_i$ not used in the specific fit are marked with --. The
    error-weighted average over all four inverse temperatures are listed as
    $\beta^{\rm av}$, whereas $\beta^{\rm av}_\text{w/o eqw}$ is the average, 
    where $\beta^{\rm eqw}$ is left out, because it is strongly correlated {with}
    $\beta^{C^{\rm max}_V}$, and would effectively weight this value twice. 
  }
  \vspace{3ex} 
  \begin{tabular}{rc*{8}{l}} 
  \toprule
  \multicolumn{1}{c}{${\cal O}$} & 
  \multicolumn{1}{c}{Eq.} & 
  \multicolumn{1}{c}{$L_{\rm min}$} & 
  \multicolumn{1}{c}{$\beta^\infty$} & 
  \multicolumn{1}{c}{$p_2$} & 
  \multicolumn{1}{c}{$p_3$} & 
  \multicolumn{1}{c}{$p_4$} & 
  \multicolumn{1}{c}{$Q$} & 
  \multicolumn{1}{c}{dof} \\
  \midrule 
  \multicolumn{9}{c}{reduced time series, linear fit}\\
  $\beta^{C_V^{\rm max}}(L)$ & \eref{eq:fss:beta:specheat2} & 16 & 0.551221(11) & $  -2.281(5)$ & -- & -- & 0.54 & 10 \\ 
  $\beta^{B^{\rm min}}(L)  $ & \eref{eq:fss:beta:binder2}   & 13 &  0.551347(7) & $-2.4373(24)$ & -- & -- & 0.79 & 13 \\
  $\beta^{\rm eqw}(L)      $ & \eref{eq:fss:beta:eqw2}      & 16 & 0.551221(11) & $  -2.281(5)$ & -- & -- & 0.54 & 10 \\
  $\beta^{\rm eqh}(L)      $ & \eref{eq:fss:beta:eqw2}      & \hphantom{1}5 & 0.551331(21) & $  -2.366(6)$ & -- & -- & 0.96 & 21 \\
  \\
  \multicolumn{2}{r}{$\beta^{\rm av}         $     }   &  & 0.551291(7) & 1.313527(12) \\
  \multicolumn{2}{r}{$\beta^{\rm av}_\text{w/o eqw}$ } &  & 0.551311(7) & 1.313493(12) \\
  \\
  \multicolumn{9}{c}{full time series, linear fit}\\
  $\beta^{C_V^{\rm max}}(L)$ & \eref{eq:fss:beta:specheat2} & 17 & 0.551233(10) & $   -2.287(5)$& -- & -- & 0.54 & \hphantom{1}9 \\
  $\beta^{B^{\rm min}}(L)  $ & \eref{eq:fss:beta:binder2}   & 13 & 0.551350(6)  & $-2.4389(19) $& -- & -- & 0.72 & 13 \\
  $\beta^{\rm eqw}(L)      $ & \eref{eq:fss:beta:eqw2}      & 17 & 0.551233(10) & $   -2.287(5)$& -- & -- & 0.54 & \hphantom{1}9 \\
  $\beta^{\rm eqh}(L)      $ & \eref{eq:fss:beta:eqw2}      & 12 & 0.551277(5)  & $-2.3478(16) $& -- & -- & 0.90 & 14 \\
  \\
  \multicolumn{2}{r}{$\beta^{\rm av}         $     }   &  & 0.551293(5) & 1.313524(9) \\
  \multicolumn{2}{r}{$\beta^{\rm av}_\text{w/o eqw}$ } &  & 0.551300(5) & 1.313511(9) \\ 
  \\
  \multicolumn{9}{c}{full time series, up to ${\cal O}\left( L^{-3} \right)$}\\
  $\beta^{B^{\rm min}}(L)  $ & \eref{eq:fss:beta:binder2}   & 11 & 0.551403(14) & $-2.494(11)$  & 0.65(12) & -- & 0.59 & 14 \\
  $\beta^{\rm eqh}(L)      $ & \eref{eq:fss:beta:eqw2}      & 12 & 0.551271(17) & $-2.342(15)$  & 0(0.2)   & -- & 0.87 & 13 \\ 
  \\
  \multicolumn{2}{r}{$\beta^{\rm av}         $     }   &  & 0.551269(10) & 1.313565(17) \\
  \multicolumn{2}{r}{$\beta^{\rm av}_\text{w/o eqw}$ } &  & 0.551288(10) & 1.313532(17) \\
  \\
  \multicolumn{9}{c}{full time series, up to ${\cal O}\left( L^{-4} \right)$}\\
  $\beta^{C_V^{\rm max}}(L)$ & \eref{eq:fss:beta:specheat2} & \hphantom{1}9 & 0.551331(8)  & $-2.371(4)$  & --        & $\hphantom{-}16.9(4)$   & 0.95 & 16 \\ 
  $\beta^{B^{\rm min}}(L)  $ & \eref{eq:fss:beta:binder2}   &  \hphantom{1}9 & 0.551340(28) & $-2.39(4)$   & $-1.6(8)$ & $\hphantom{-}13(4)$     & 0.92 & 15 \\
  $\beta^{\rm eqw}(L)      $ & \eref{eq:fss:beta:eqw2}      &  \hphantom{1}9 & 0.551331(8)  & $-2.371(4)$  & --        & $\hphantom{-}17.0(4)$   & 0.95 & 16 \\
  $\beta^{\rm eqh}(L)      $ & \eref{eq:fss:beta:eqw2}      & 12 & 0.55134(6)   & $-2.47(11)$  & $\hphantom{-}2.8(2.4)$  & $-18(15)$ & 0.95 & 12 \\ 
      \\
  \multicolumn{2}{r}{$\beta^{\rm av}         $     }   &  & 0.551332(8) & 1.313457(14) \\
  \multicolumn{2}{r}{$\beta^{\rm av}_\text{w/o eqw}$ } &  & 0.551332(8) & 1.313456(14) \\ 
  \bottomrule
\end{tabular} 

  \label{tab:resultingfits}
\end{table} 

\clearpage

Since the inverse temperatures $\beta^{\rm eqw}$ and $\beta^{C_V^{\rm max}}$
are obviously strongly correlated, we leave out the former and average
over $\beta^{C_V^{\rm max}}, \beta^{B^{\rm min}}$, and $\beta^{\rm eqh}$,
neglecting cross-correlations~\cite{combinefitsweigel} between those.
Our best estimate {of} the inverse transition temperature is then given by
\begin{eqnarray}
  \beta^\infty &= 0.551\,332(8) &\qquad\textrm{original model, periodic bc,}
\end{eqnarray} 
which accounts for the higher-order scaling corrections up to 
${\cal O}\left(L^{-4}\right)$. 

Although the inverse transition temperatures do not change, we employ the full
data set. The reason is that the error on $\beta^{\rm eqh}$ becomes smaller
for the time series that uses the full, correlated data set. This is attributed
to the fact that the observable relies on the statistics in single bins of the
energy histogram, which in total becomes smoother when using more, correlated
measurements.  The same argument is valid for the calculation of the interface
tension{~\eqref{eq:interface-tension}, for which the best fit 
with corrections of order ${\cal O}(L^{-2})$ yields a value of}
\begin{eqnarray}
  \hat\sigma = 0.12037(18) &\qquad\textrm{original model, periodic bc.}
\end{eqnarray}

Moments of the energy in the pure ordered and disordered phases are
also expected to become more accurate using the full data set, since
autocorrelation times in the pure phases are {then} significantly smaller
{than $\tau^{\rm int}$ for the full energy range (see below).
By fitting the scaling law \eref{eq:fss:energies} to these moments, one obtains the latent
heat in the infinite-volume limit,}
\begin{eqnarray}
  \Delta\hat{e} = 0.85148(5) &\qquad\textrm{original model, periodic bc.}
\end{eqnarray}

Taking a careful look at the scaling laws in section~\ref{sec:fss}, we find
that the prefactors of the scaling corrections only depend on the moments of
the energy or their differences. We have two methods at hand to test the
self-consistency of our simulations. {Firstly, since} the statistics of the observables
{are} very high, we can retrieve the prefactors of the corrections as
parameters of (nonlinear) least-squares-fits with all corrections up to and
including ${\cal O}(L^{-4})$. Secondly, from multicanonical simulations we get
estimators~\eref{eq:e-eqw} of the energy moments, allowing a direct 
computation of those prefactors.

{In addition}, we {carried out} independent canonical simulations for the original model
under periodic boundary conditions for very large lattices. The goal was to get 
independent measurements {of} the moments of the energy in the ordered and
disordered phases. Here we prepared the system in the appropriate phase and
performed the simulations at a fixed temperature $\beta = 0.5513$, near the
transition temperature, exploiting {the fact} that {in canonical simulations}, for large lattices, flips between the
two phases are extremely unlikely.  
Of course, {this} was only possible after having determined the transition 
temperature with high accuracy by the multicanonical simulations. 

The quality of the canonical measurements and estimators on the energy and the
specific heat are summarized in Table~\ref{tab:canonicalresults}.  The
autocorrelation times within the phases are significantly smaller, because the
system {does} not traverse suppressed, {improbable} states between the phases.  The
statistical error has  again been retrieved by jackknife analysis.  For
lattices with size $L\geq 32$, physical quantities indicate no further
dependence on the lattice size within the error. Therefore we can safely take
the error-weighted averages over energy moments and their differences for those
lattices. The multicanonical and the canonical estimates {of} energetic moments
agree astonishingly well.

With use of the energy moments {from} both simulations, we can challenge the 
prefactors {in} {the finite-size scaling laws}
numerically by comparing the numerical values for the fit parameters
to the {theoretically} expected prefactors {in terms of the energy moments.}
The results of this cross-check {are collected} in Table~\ref{tab:consistency}.

\begin{table}[htpb]
  \tablesize 
  \centering 
  \caption{Quality and resulting quantities of the canonical simulations. In the time series of $n_{\rm meas} = 1048576$ measurements we found autocorrelation times $\tau < 7$ (in units of measurements), leading to approximately $n$ uncorrelated measurements. The energy $\hat{e}_{\rm o/d}$ and the specific heat $\hat{C}_{\rm o/d}$ in the different phases have been measured by preparing two independent systems for each respective phase at inverse temperature $\beta = 0.5513$. We give the error-weighted average over lattices $L\geq32$, where the dependence on the lattice size is smaller than the statistical error. The last line gives infinite-volume limits from the multicanonical simulations for comparison.}
  \vspace{3ex}
  \begin{tabular}{l*{8}{l}} 
    \toprule
    \multicolumn{1}{c}{$L$} &
    \multicolumn{1}{c}{$n$} & 
    \multicolumn{1}{c}{$\hat{e}_o$} &
    \multicolumn{1}{c}{$\hat{e}_d$} &
    \multicolumn{1}{c}{$\Delta \hat{e}$} &
    \multicolumn{1}{c}{$\hat{C}_o$} &
    \multicolumn{1}{c}{$\hat{C}_d$} &
    \multicolumn{1}{c}{$\Delta \hat{C}$}
    \\
    \midrule
    16 & 149796 & $-1.468406(29)$ & $-0.61804(13) $ & 0.85036(21)  & 0.1645(5) & 0.861(4)   & 0.696(6)\\
    32 & 174762 & $-1.468362(10)$ & $-0.61741(4)  $ & 0.85095(6)   & 0.1645(6) & 0.8464(29) & 0.682(5)\\
    48 & 149796 & $-1.468360(6) $ & $-0.617382(19)$ & 0.850978(30) & 0.1658(5) & 0.8445(27) & 0.679(5)\\
    64 & 174762 & $-1.468367(6) $ & $-0.617401(15)$ & 0.850966(24) & 0.1656(6) &   0.847(4) & 0.681(6)\\
    \midrule                                                                                                 
  \multicolumn{2}{l}{average $(L\geq 32)$} & $-1.468364(4)$ & $-0.617396(11)$ & 0.850968(18) & 0.16534(30) & 0.8458(18) & 0.6805(27) \\
  \multicolumn{2}{l}{multicanonical} & $-1.468373(12)$ & $-0.61771(6)$   & 0.85148(5)   & 0.16414(15) & 0.8410(12) & 0.6769(17)\\
    \bottomrule
    \end{tabular}
    \label{tab:canonicalresults} 
\end{table} 

\clearpage

Employing the scaling relation for the specific-heat
maximum~\eref{eq:fss:specheat}, we can calculate $\Delta\hat{e}$ from the fit
parameter of the leading contribution. Using our estimate {of} $\beta^{\infty} =
0.551\,332(8)$, we get $\Delta\hat{e} = 0.85130(7)$, very close to our estimate
$0.850968(18)$ from the moments of the canonical simulations.
The leading correction to the specific-heat ansatz~\eref{eq:fss:specheat} has a
prefactor which computes to {$0.2197(17)$ from the canonical moments}.
The fits find $0.17(6)$, which is compatible, {if} not quite accurate.

The minimum of the 
Binder parameter~\eref{eq:fss:binder} for the infinite lattice is
found to be $0.34729(7)$ from the direct fit {of} our multicanonical data which
{agrees} within error bars {with} $0.34723(9)$ from the canonical and $0.347(4)$
from the multicanonical energy moments. The first correction {in}
Eq.~\eref{eq:fss:binder:a2} yields a value of $-9.195(14)$ when inserting the
energy {moments} from the multicanonical simulation. The fits find $-9.12(4)$
which is very close.

\begin{table}
  \tablesize 
  \centering 
  \caption{Resulting prefactors of the finite-size scaling corrections of the original model, retrieved by fitting the ansatz, compared to direct calculations from estimators for the energy $\hat{e}_{\rm o/d}$ and specific heat $\hat{C}_{\rm o/d}$ of the ordered and disordered phases. In the multicanonical simulations these moments were determined by finite-size scaling of $e_{\rm o/d}(L), c_{\rm o/d}(L)$; and in the canonical case by measuring time series directly at $\beta = 0.5513\simeq\beta^\infty$.}
  \vspace{3ex} 
  \begin{tabular}{*{6}{l}} 
  \toprule
  input & 
  $\Delta\hat{e}$ & 
  $\frac{3\ln(2)}{\Delta \hat{e}}$ & 
  $\frac{2\ln(\hat{e}_{\rm o}/\hat{e}_{\rm d})}{\Delta \hat{e}}$ &
  $\frac{\ln(\hat{C}_{\rm d}/\hat{C}_{\rm o})}{2\Delta \hat{e}}$ &
  $B^{\rm min}_{L\rightarrow\infty}$\\
  \midrule
  fit on $C^{\rm max}_{V}(L)$                                 & 0.85130(7)   & --           & --          & --         & --\\
  fit on $B^{\rm min}(L)$                                     & --           & --           & --          & --         & 0.34729(7) \\
  fit on $\beta^{C^{\rm max}_{V}}(L)$                         & 0.8771(14)   & $2.371(4)$   & --          & --         & --\\
  fit on $\beta^{B^{\rm min}}(L)$                             & 0.871(14)    & $2.39(4)$    & 1.6(8)      & --         & --\\
  fit on $\beta^{\rm eqw}(L)$                                 & 0.8770(14)   & $2.371(4)$   & --          & --         & --\\
  fit on $\beta^{\rm eqh}(L)$                                 & 0.84(4)      & $2.47(11)$   & --          & 2.8(2.4)   & --\\
  fit on $\beta^{C^{\rm max}_{V}} - \beta^{B^{\rm min}}$      & --           & --           & 2.03469(7)  & --         & --\\
  fit on $\beta^{C^{\rm max}_{V}} - \beta^{\rm eqh}$          & --           & --           & --          & {0.892(14)}  & --\\
  \midrule
  \multicolumn{6}{l}{energy moments from simulations \dots}\\
  \multicolumn{1}{r}{\dots multicanonical}
                                               & 0.85148(5)   & $2.44215(15)$ & 2.03649(27) & 0.9594(10) & 0.347(4)   \\
  \multicolumn{1}{r}{\dots canonical}
                                               & 0.850968(18) & $2.44362(6)$ & 2.03625(6)  & 0.9591(16) & 0.34723(9) \\
  \bottomrule 
  \end{tabular} \label{tab:consistency}
\end{table}

The coefficient of the leading correction apparent for all inverse temperatures, 
$p_2 = 3\ln(2)/\Delta \hat{e}$, agrees reasonably well for the
fits on $\beta^{B^{\rm min}}$ and $\beta^{\rm eqh}$ {(cf.~Table~\ref{tab:consistency})}. The fits on 
$\beta^{\rm eqw}$ and $\beta^{C^{\rm max}_V}$ yield a slope of $2.371({4})$ which 
within error bars is slightly off {from the value of} our best estimate of $2.44362(6)$ {from the energy moments}. 
The relative error between the two values is very small though, around $3\%$,
which is acceptable given that the leading contribution probably accounts for
the omitted higher-order contributions with different sign and that we {have}
neglected all exponential corrections.

The second leading correction of order ${\cal O}(L^{-3})$ of $\beta^{B^{\rm
min}}$ has a prefactor of the form $2\ln(\hat{e}_{\rm o}/\hat{e}_{\rm d})/\Delta \hat{e}$, which
we expect to have a value of $2.03625(6)$ from the energy moments.  The
corrections of fourth order {to} $\beta^{B^{\rm min}}, \beta^{C^{\rm max}_V}$ and
$\beta^{\rm eqw}$ are supposed to be identical from the analytical expansion in
section~\ref{sec:fss}. The fits of the inverse temperatures $\beta^{C^{\rm
max}_V}$ and $\beta^{\rm eqw}$ in Table~\ref{tab:resultingfits} suggest a value
around $17$ for {the ${\cal O}(L^{-4})$ contribution}. For the lattice sizes
accessible to the multicanonical approach, that is of the same absolute order of
magnitude (but different sign) compared to the third-order contribution. Therefore
they should, in principle, compensate each other. This is reflected by the fact
that the second-order contribution $p_2$ of $\beta^{B^{\rm min}}$ is closest to
the expected one. In accordance, fitting the observable to the law
$\beta^{\infty} + p_2/L^2 + p_3/L^3$ gives a prefactor $p_3 = 0.65(12)$ with
the wrong sign compared to the theoretical prediction~\eref{eq:fss:beta:binder2},
compensating the next contribution.  We therefore also {looked} at the fit
including the fourth term of order ${\cal O}(L^{-4})$.  Not taking the {explicit values}
too {seriously}, we find $p_3 = -1.6(8),\, p_4 = 13(4)$, which reflect qualitatively
the compensation of those contributions for our lattice sizes at hand. Overall,
we must conclude that least-squares fitting {cannot be pushed any further} given our
statistics.

The observation that $\beta^{C^{\rm max}_V}$ and $\beta^{B^{\rm min}}$ have the
same ${\cal O}(L^{-4})$-contribution can be exploited (implicitly also making
use of the cross-correlations) by looking at their difference.
Here, we expect from \eref{eq:fss:beta:specheat2} and
\eref{eq:fss:beta:binder2} a remainder of $2\ln(\hat{e}_{\rm o}/\hat{e}_{\rm d})/L^3\Delta
\hat{e} + {\cal O}(L^{-5})$ for the scaling. In fact, fitting the difference gives a
prefactor $2.03469(7)$, in {excellent} agreement with $2.03625(6)$ {from the formula}, where the
relative error between the two is less than $0.1\%$.

The difference $\beta^{C^{\rm max}_{V}} - \beta^{\rm eqh}$ should give the
third correction {to} $\beta^{\rm eqh}$, {which} reads $p_3 = \ln(\hat{C}_{\rm
d}/\hat{C}_{\rm o})/2\Delta \hat{e}$. The fit yields $0.892(14)$ with $Q=0.98$
and $8$ degrees of freedom left, which differs from {$0.9594(10)$} by about
$7\%$.

Finally, we can also compare the numerical values for the correction to the
energies via~\eref{eq:fss:energies}. For $\hat{e}_{\rm o}$, we find a prefactor
of $1.329(5)$ from the specific heat $\hat{C}_{\rm o}$, compared to the value
of $1.397(5)$, for $\hat{e}_{\rm d}$ a value of $6.80(3)$ compared to
$6.09(4)$, which is roughly $10\%$ off.

The overall consistency of our results is very good, given that we neglected
all exponential corrections. No estimates for the prefactors differ by more than
$10\%$, and the {various}
estimates of the inverse transition temperature are insensitive to the actual 
fitting protocol we use.  This clearly demonstates that the first
correction terms are properly predicted by the simple two-state model even in
the case of models with an exponential degeneracy of the low-temperature phase.

%
The earlier canonical Monte Carlo simulations of the original plaquette model
yielded values of $\beta^\infty = 0.50(1)$~\cite{Johnston1996} and more
recently canonical simulations of the  dual model~\eref{eq:ham:dual}  gave
$\beta^\infty = 0.510(2)$~\cite{Johnston2011}.
Another previous estimate for the infinite-lattice inverse transition
temperature, reported by Baig et al.~\cite{Baig2004} from canonical simulations
using fixed boundary conditions, $\beta^\infty = 0.54757(63)$, is much closer
to the results here.

We therefore measured the inverse transition temperature again using
multicanonical simulations for both the dual model~\eref{eq:ham:dual} under 
periodic boundary conditions and the original model~\eref{eq:ham:gonikappa0} 
with fixed boundary conditions. We resolve those inconsistencies, as we  
show in the following.

\subsection{Dual model with periodic boundary conditions}

\begin{figure}[h]
  \begin{center} 
    \includegraphics[scale=0.98]{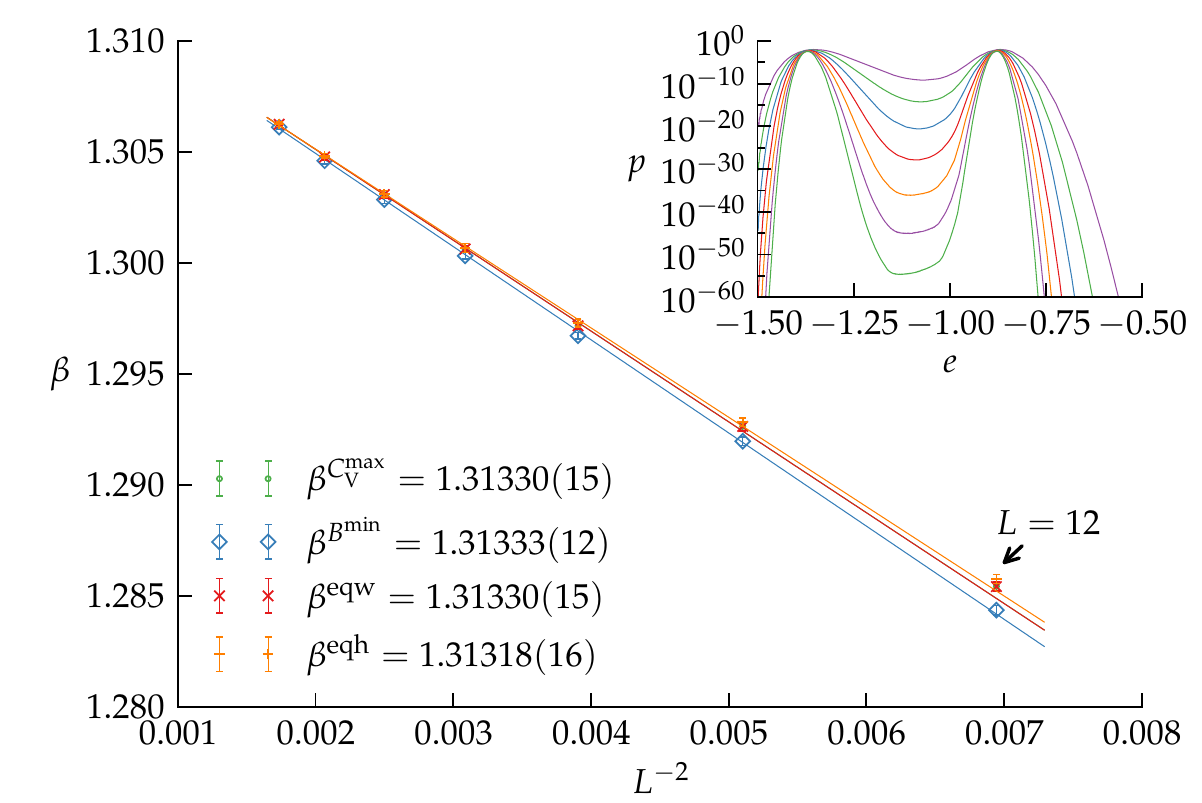} 
    \caption{Best
      fits obtained for the dual model with periodic boundary conditions using
      the (finite lattice) peak locations for the specific heat $C_V^{\rm
      max}$, Binder's energy parameter $B^{\rm min}$; or inverse temperatures
      $\beta^{\rm eqw}$ and $\beta^{\rm eqh}$, where the two peaks of the
      energy probability density are of same weight or have equal height,
      respectively.
      The inset shows the energy probability density $p(e)$ over $e = E/L^d$ at
      $\beta^{\rm eqh}$
      {for lattices of linear size $L\in \left\{ 12, 14, \dots, 24 \right\}$.} } 
    \label{fig:fit:dual} 
  \end{center} 
\end{figure}

For the dual model, we performed a number of $n^{\rm max} = 4\times 10^6$ sweeps
and took measurements every sweep for even lattices up to $L=24$.
The inverse temperatures of the dual model were fitted to laws with the 
leading correction of order ${\cal O}(L^{-2})$, which should be well covered by 
our data. The best fits on the inverse temperatures are shown in
Figure~\ref{fig:fit:dual}, where we used the data in
Table~\ref{tab:dualresults} that also lists the other quantities of interest.

\begin{landscape} \begin{table}[htpb] \tablesize \centering 
\caption{ Simulation results for the dual model~\eref{eq:ham:dual}: extremal
      values for the specific heat $C_V^{\rm max}$, Binder's energy parameter
      $B^{\rm min}$, with their respective pseudo-critical inverse temperatures
      $\beta$, and temperatures where peaks of the energy probability density
      have equal heights and weights for finite lattices with linear length
      $L$. The finite-lattice interface tension is listed as $\sigma$, the
      energy of the ordered and disordered phases and their difference as
      $e_{\rm o}, e_{\rm d}$ and $\Delta e$. Light grey cells mark the values
      used for fitting, so that the goodness-of-fit parameter $Q>0.5$. If $Q <
      0.5$ for all fits, we took that one with the largest $Q$.}
      \vspace{3ex}
  \begin{tabular}{l*{10}{r}} \toprule \multicolumn{1}{c}{$L$} &
    \multicolumn{1}{c}{$\beta^{C_V^{\rm max}}$} & \multicolumn{1}{c}{$C_V^{\rm
    max}$} & \multicolumn{1}{c}{$\beta^{B^{\rm min}}$} &
    \multicolumn{1}{c}{$B^{\rm min}$} & \multicolumn{1}{c}{$\beta^{\rm eqw}$} &
    \multicolumn{1}{c}{$\beta^{\rm eqh}$} & \multicolumn{1}{c}{$\sigma$} &
    \multicolumn{1}{c}{$e_{\rm o}$} & \multicolumn{1}{c}{$e_{\rm d}$} &
    \multicolumn{1}{c}{$\Delta e$} \\ \midrule
 08 & 1.25788(22)          & 37.26(11)            & 1.25394(22)          & 0.61158(15)        & 1.2577(11)           & 1.25891(28) & 0.0232(4)  & $-1.303(26)  $ & $-0.891(21)  $ & 0.412(6)              \\
 10 & 1.27493(21)         & 89.34(12)            & 1.27304(21)   & 0.60056(9)         &  1.27488(21)   &  1.27557(29) & 0.0412(5)  & $-1.3329(4)  $ & $-0.86898(28)$ & 0.46395(30) \\
 12 & 1.28544(22)         & 165.62(25)           & \hllg{1.28437(22)}   & 0.59735(10)        &  1.28543(22)   &  1.28577(21) & 0.0574(7)  & $-1.3476(4)  $ & $-0.86823(24)$ & 0.4794(4)             \\
 14 & \hllg{1.29265(20)}   & 271.57(30)           & \hllg{1.29198(20)}   & 0.59642(7)         & \hllg{1.29264(20)}   & \hllg{1.29283(20)} & \hllg{0.0715(7)}  & \hllg{$-1.35585(30)$} & $-0.87042(19)$ & 0.48544(25)           \\
 16 & \hllg{1.29717(15)}   & \hllg{412.9(4)}      & \hllg{1.29673(15)}   & \hllg{0.59601(7)}  & \hllg{1.29717(15)}   & \hllg{1.29730(17)} & \hllg{0.0829(7)}  & \hllg{$-1.36106(18)$} & \hllg{$-0.87238(14)$} & \hllg{0.48868(22)} \\
 18 & \hllg{1.30063(16)}   & \hllg{592.6(5)}      & \hllg{1.30032(16)}   & \hllg{0.59616(5)}  & \hllg{1.30063(16)}   & \hllg{1.30068(21)} & \hllg{0.0911(7)}  & \hllg{$-1.36412(17)$} & \hllg{$-0.87451(13)$} & \hllg{0.48961(16)}           \\
 20 & \hllg{1.30308(18)}   & \hllg{819.0(5)}      & \hllg{1.30286(18)}   & \hllg{0.59613(6)}  & \hllg{1.30308(18)}   & \hllg{1.30312(15)} & \hllg{0.0974(13)} & \hllg{$-1.36657(13)$} & \hllg{$-0.87589(15)$} & \hllg{0.49067(18)}           \\
 22 & \hllg{1.30478(15)}   & \hllg{1095.5(9)}     & \hllg{1.30461(15)}   & \hllg{0.59613(4)}  & \hllg{1.30478(15)}   & \hllg{1.30483(11)} & \hllg{0.101(4)}   & \hllg{$-1.36826(20)$} & \hllg{$-0.87691(13)$} & \hllg{0.49136(16)}           \\
 24 & \hllg{1.30625(9) }   & \hllg{1425.8(9)}     & \hllg{1.30612(9)}    & \hllg{0.59626(5)}  & \hllg{1.30625(9)}    & \hllg{1.30626(11)} & \hllg{0.1044(11)} & \hllg{$-1.36945(14)$} & \hllg{$-0.87796(11)$} & \hllg{0.49149(14)}           \\
\midrule 
\rowcolor{lightgrey} $\infty$ & 1.31330(15)   & 0.10511(12)$L^3$  &  1.31333(12)         & 0.59636(8)  & 1.31330(15)          & 1.31318(16)          & 0.1214(13)           & $-1.37644(21)$         & $-0.88227(19)$         & 0.49402(26)  \\
\rowcolor{lightgrey} $Q$ & 0.58          & 0.48              &  0.58                & 0.18        & 0.52                 & 0.66                      & 0.99                 & 0.63                   & 0.54                 & 0.27                  \\

    \bottomrule 
    \end{tabular}\label{tab:dualresults}
\end{table} \end{landscape}

Since the inverse temperatures $\beta^{\rm eqw}$ and $\beta^{C_V^{\rm max}}$
are again obviously strongly correlated, we leave out the former and average
over $\beta^{C_V^{\rm max}}, \beta^{B^{\rm min}}$, and $\beta^{\rm eqh}$,
neglecting cross-correlations~\cite{combinefitsweigel} between those.
We then find the error weighted averages,
\begin{eqnarray}
  \beta^\infty_{\rm dual} &= 1.313\,28(12) &\qquad\textrm{dual model, periodic bc,} 
\end{eqnarray} 
for the inverse transition temperatures of the models. The error is taken 
as the smallest error of the contributing estimates. 

The temperature $\beta^\infty_{\rm dual}$ of the dual model is related to the temperature in
the original model, $\beta^\infty$, by the duality transformation 

\begin{equation}
  \beta^\infty = -\ln\left( \tanh\left( \frac{\beta^\infty_{\text{dual}}}{2}
  \right) \right). \label{eq:duality} 
\end{equation} 
Applying standard error propagation, we retrieve a value of 
\begin{eqnarray} 
  \beta^\infty &=
  0.551\,43(7) \qquad &\textrm{from duality, periodic bc} 
\end{eqnarray} 
for the original model. This agrees very well with $0.551\,332(8)$, {obtained}
from the direct simulation, considering that higher-order corrections in the
dual model are omitted and additional exponential
{corrections~\cite{rigorous-fss-potts,twostatemodel,exponentialcorr}} in the
finite-size scaling were ignored completely for both models.

\begin{figure}[htb] \begin{center} \includegraphics[]{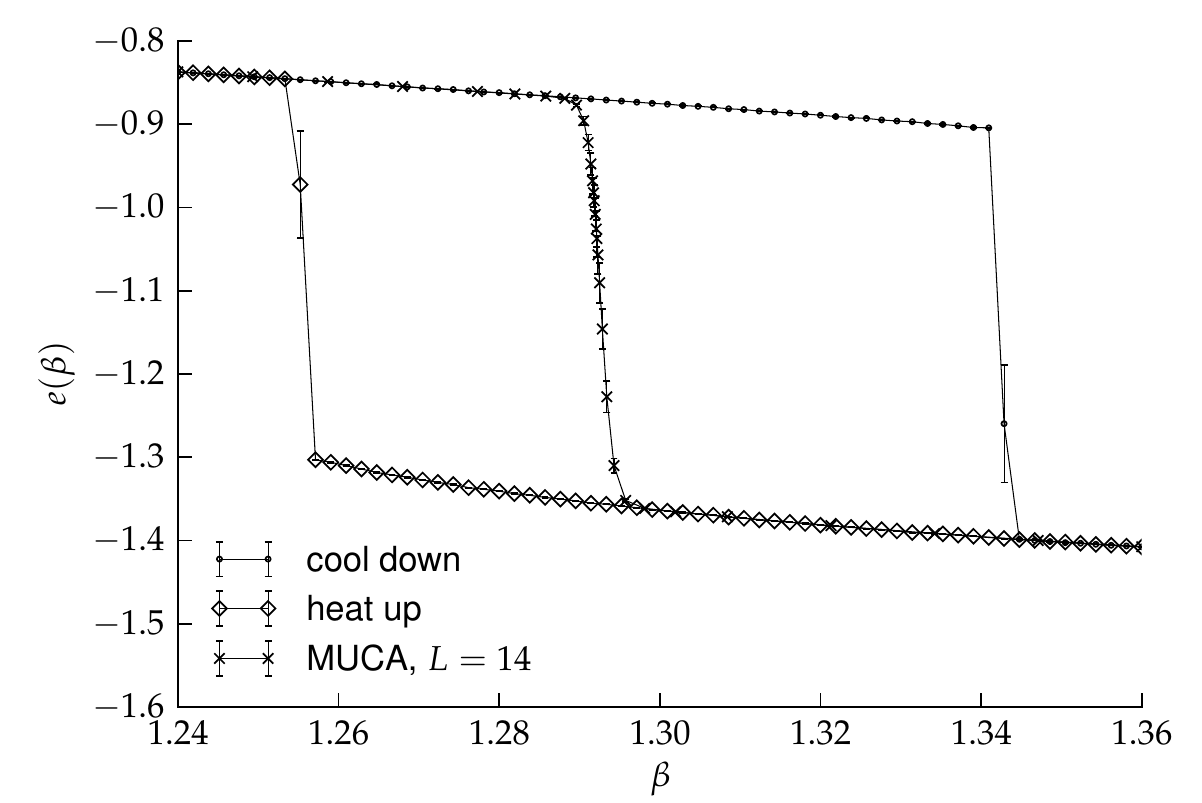}
    \caption{Strong hysteresis effects in the dual gonihedric
      Ising model with periodic boundary conditions. The linear lattice size is
      $L=14$ for comparison with Fig.~10 of Ref.~\cite{Johnston2011}. One can
      see that heating the system up (decreasing the inverse temperature
      $\beta$) or cooling it down (increasing $\beta$) can lead to strong
      hysteresis effects in the energy. Our multicanonical data lies in between
      both branches of the hysteresis curve, but not in the centre, as one may
      heuristically assume.}
\label{fig:hysteresis} \end{center} \end{figure}

We argue that the earlier estimates on the transition temperature were clearly
hampered by strong hysteresis effects. Apart from the locations of the
hysteresis branches being cooling-rate dependent, it is hard to estimate the
transition temperature reliably from their locations. This is illustrated in
Figure~\ref{fig:hysteresis}, where the multicanonical data of the dual model is
located between the two hysteresis branches.
Such effects are very {difficult} to tackle using canonical
Monte Carlo data, as already remarked on by the authors of
Ref.~\cite{Johnston2011}. 

{
For the interface tension~\eref{eq:interface-tension} of the dual model we find
$\sigma = 0.1214(13)$. This value agrees very well with the interface tension of 
the original model, $\sigma = 0.12037(18)$, which raises interesting questions 
about the duality of the model.}

\subsection{Original plaquette model with fixed boundary conditions} 

The remaining open question is the difference between our {inverse transition temperature} compared to the
value {in} the same model under fixed boundary conditions. In the thermodynamic
limit, we expect a system to be independent of its boundary conditions, since
the boundaries grow like a surface, whereas the system size grows with the
volume.  
We therefore reinvestigated the model~\eref{eq:ham:gonikappa0} using the multicanonical
algorithm for such fixed boundary conditions.

\begin{landscape} \begin{table}[htpb] \tablesize \centering 
      \caption{Extremal points of the gonihedric Ising
      model~\eref{eq:ham:gonikappa0} for the specific heat $C_V^{\rm
      max}$, Binder's energy parameter $B^{\rm min}$, with their respective
      pseudo-critical inverse temperatures $\beta$ for finite lattices with
      linear length $L$ contained in a box with \emph{fixed} boundary conditions. For
      each lattice size, a number of $n_{\rm prod} = 10^6$ measurements were
    taken. Errors have been calculated with jackknife error estimation using
  $20$ blocks.} \vspace{3ex} \begin{tabular}{l*{10}{r}} \toprule
    \multicolumn{1}{c}{$L$} & \multicolumn{1}{c}{$\beta^{C_V^{\rm max}}$} &
    \multicolumn{1}{c}{$C_V^{\rm max}$} & \multicolumn{1}{c}{$\beta^{B^{\rm
    min}}$} & \multicolumn{1}{c}{$B^{\rm min}$} &
    \multicolumn{1}{c}{$\beta^{\rm eqw}$} & \multicolumn{1}{c}{$\beta^{\rm
    eqh}$} & \multicolumn{1}{c}{$\sigma$} & \multicolumn{1}{c}{$e_{\rm o}$} &
    \multicolumn{1}{c}{$e_{\rm d}$} & \multicolumn{1}{c}{$\Delta e$} \\
    \midrule
 08 & 0.44699(14)          & 5.29(4)              & 0.44233(13)          & 0.63030(26)         & 0.4475(18)            & 0.4479(8)            & 0(0.002)             & $-1.67(5)     $        & $-1.29(6)   $               & 0.387(7)              \\
 09 & 0.45794(13)          & 7.63(8)              & 0.45488(14)          & 0.6305(4)           & 0.4594(12)            & 0.4577(11)           & 0(0.003)             & $-1.65(4)     $        & $-1.28(5)   $               & 0.379(10)             \\
 10 & 0.46715(15)          & 11.08(12)            & 0.46510(16)          & 0.6293(5)           & 0.4678(6)             & 0.46683(26)          & 0(0.003)             & $-1.623(21)   $        & $-1.230(29) $               & 0.394(9) \\
 11 & 0.47465(14)          & 15.81(14)            & \hllg{0.47323(14)}   & 0.6273(4)           & 0.4752(5)             & 0.47403(25)          & 0.0044(8)            & $-1.62(4)     $        & $-1.21(6)   $               & 0.410(20) \\
 12 & 0.48097(14)          & 22.70(26)            & \hllg{0.47995(14)}   & 0.6234(6)           & 0.48136(19)           & 0.4803(4)            & 0.0060(13)           & $-1.625(9)    $        & $-1.192(19) $               & 0.433(11)             \\
 13 & 0.48629(9)           & \hllg{31.56(30)}     & \hllg{0.48552(9)}    & 0.6197(6)           & 0.48653(9)            & 0.48566(9)           & 0.0072(14)           & $-1.6230(7)   $        & $-1.1654(25)$               & 0.4575(27) \\
 14 & 0.49086(8)           & \hllg{43.7(5)}       & \hllg{0.49027(9)}    & 0.6146(7)           & 0.49099(11)           & \hllg{0.49020(27)}   & 0.0090(7)            & $-1.620(8)    $        & \hllg{$-1.129(17) $}        & 0.490(9)              \\
 15 & 0.49483(13)          & \hllg{56.8(8)}       & \hllg{0.49435(13)}   & 0.6116(9)           & 0.49490(17)           & \hllg{0.49429(12)}   & 0.0097(11)           & $-1.612(14)   $        & \hllg{$-1.104(28) $}        & \hllg{0.508(14)} \\
 16 & \hllg{0.49825(11)}   & \hllg{75.2(8)}       & \hllg{0.49786(11)}   & 0.6059(8)           & 0.49830(11)           & \hllg{0.49778(10)}   & 0.0119(4)            & $-1.6108(5)   $        & \hllg{$-1.080(4)  $}        & \hllg{0.531(4)} \\
 17 & \hllg{0.50126(7)}    & \hllg{92.6(8)}       & \hllg{0.50093(7)}    & 0.6043(7)           & \hllg{0.50129(7)}     & \hllg{0.50090(19)}   & \hllg{0.0125(20)}    & $-1.6046(5)   $        & \hllg{$-1.0664(27)$}        & \hllg{0.5382(25)}            \\
 18 & \hllg{0.50410(6)}    & \hllg{115.3(11)}     & \hllg{0.50383(6)}    & \hllg{0.6008(8)}    & \hllg{0.50412(6)}     & \hllg{0.50381(24)}   & \hllg{0.0135(19)}    & $-1.59885(24) $        & \hllg{$-1.0487(28)$}        & \hllg{0.5501(28)}            \\
 19 & \hllg{0.50648(8)}    & \hllg{141.9(12)}     & \hllg{0.50625(8)}    & \hllg{0.5969(8)}    & \hllg{0.50649(8)}     & \hllg{0.50620(14)}   & \hllg{0.0147(15)}    & $-1.59286(27) $        & \hllg{$-1.0311(24)$}        & \hllg{0.5618(24)}            \\
 20 & \hllg{0.50866(10)}   & \hllg{167.3(16)}     & \hllg{0.50846(10)}   & \hllg{0.5963(10)}   & \hllg{0.50867(10)}    & \hllg{0.50841(12)}   & \hllg{0.0156(13)}    & $-1.58807(27) $        & \hllg{$-1.0248(30)$}        & \hllg{0.5633(29)}            \\
 21 & \hllg{0.51063(8)}    & \hllg{200.7(19)}     & \hllg{0.51046(8)}    & \hllg{0.5930(9)}    & \hllg{0.51064(8)}     & \hllg{0.51039(7)}    & \hllg{0.01706(26)}   & $-1.5828(4)   $        & \hllg{$-1.0106(27)$}        & \hllg{0.5722(28)}            \\
 22 & \hllg{0.51258(6)}    & \hllg{235.1(21)}     & \hllg{0.51243(6)}    & \hllg{0.5914(9)}    & \hllg{0.51259(6)}     & \hllg{0.51237(5)}    & \hllg{0.0173(12)}    & $-1.57837(20) $        & \hllg{$-1.0024(26)$}        & \hllg{0.5759(26)}            \\
 23 & \hllg{0.51433(7)}    & \hllg{275.6(24)}     & \hllg{0.51420(7)}    & \hllg{0.5889(10)}   & \hllg{0.51433(7)}     & \hllg{0.51415(7)}    & \hllg{0.01905(26)}   & $-1.57402(20) $        & \hllg{$-0.9920(27)$}        & \hllg{0.5820(26)}            \\
 24 & \hllg{0.51586(6)}    & \hllg{317(4) }       & \hllg{0.51574(6)}    & \hllg{0.5879(11)}   & \hllg{0.51586(6)}     & \hllg{0.51568(6)}    & \hllg{0.0193(6)}     & $-1.57031(14) $        & \hllg{$-0.9864(28)$}        & \hllg{0.5839(28)}            \\
 25 & \hllg{0.51728(7)}    & \hllg{368.6(21)}     & \hllg{0.51718(7)}    & \hllg{0.5847(7)}    & \hllg{0.51729(7)}     & \hllg{0.51716(6)}    & \hllg{0.0200(10)}    & $-1.56641(24) $        & \hllg{$-0.9750(18)$}        & \hllg{0.5914(17)} \\
 26 & \hllg{0.51853(6)}    & \hllg{422.2(28)}     & \hllg{0.51843(6)}    & \hllg{0.5827(8)}    & \hllg{0.51853(6)}     & \hllg{0.51837(15)}   & \hllg{0.0207(11)}    & \hllg{$-1.56259(18) $} & \hllg{$-0.9670(20)$}        & \hllg{0.5956(20)} \\
 27 & \hllg{0.51985(7)}    & \hllg{472(4) }       & \hllg{0.51977(7)}    & \hllg{0.5830(8)}    & \hllg{0.51985(7)}     & \hllg{0.51971(6)}    & \hllg{0.0210(14)}    & \hllg{$-1.55980(25) $} & \hllg{$-0.9656(20)$}        & \hllg{0.5942(20)}            \\
 28 & \hllg{0.52084(6)}    & \hllg{543(5) }       & \hllg{0.52077(6)}    & \hllg{0.5795(12)}   & \hllg{0.52084(6)}     & \hllg{0.52073(9)}    & \hllg{0.0220(4)}     & \hllg{$-1.55660(11) $} & \hllg{$-0.9544(28)$}        & \hllg{0.6022(28)}            \\
 29 & \hllg{0.52198(24)}   & \hllg{603(12)}       & \hllg{0.52191(24)}   & \hllg{0.5796(23)}   & \hllg{0.52198(24)}    & \hllg{0.52190(13)}   & \hllg{0.023(5)}      & \hllg{$-1.5538(9)   $} & \hllg{$-0.953(6)  $}        & \hllg{0.601(7)}              \\
\midrule
\rowcolor{lightgrey} $\infty$ & 0.55119(11) & 0.0327(6)$L^3$ & 0.55146(7) & 0.5444(14) & 0.55119(12) & 0.55152(12) &  0.0281(7)  & $-1.4782(27)$ & $-0.790(4)$ & 0.694(4) \\
\rowcolor{lightgrey} $Q$ & 0.53        & 0.53       & 0.56       & 0.59       & 0.52        & 0.53 &  0.99 & 0.49          & 0.73        & 0.50 \\ 
\bottomrule 
\end{tabular}
\label{tab:fixedresults} \end{table} \end{landscape}

\begin{figure}[htb] \begin{center} \includegraphics[]{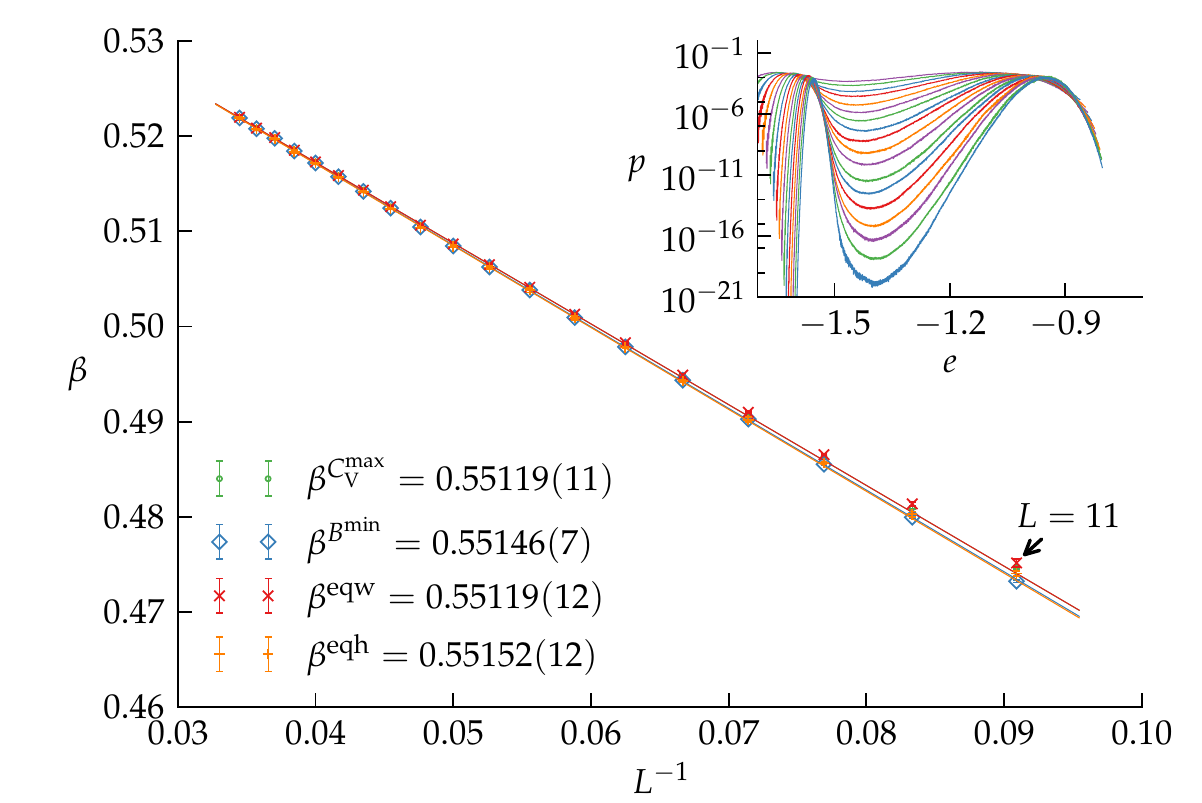} 
      \caption{
      Best fits obtained for the gonihedric Ising model using fixed boundary
      conditions using the (finite lattice) peak locations for the specific
      heat $C^{\rm max}$, Binder's energy parameter $B_{\rm min}$. The inset
      shows the energy probability density $p(e)$ over $e = E/L^3$ at
      $\beta^{\rm eqh}$ for lattices with a number of 
      {$L\in \left\{ 11, 12, \dots, 29 \right\}$}
      free spins in each dimension.} \label{fig:fit:orig-fbc}
    \end{center} \end{figure}

For our simulations we enclosed $L^3$ free spins in a larger cube with frozen
outer planes, so the whole system contained $(L+2)^3$ spins. Our results are
listed in Table~\ref{tab:fixedresults}.  We performed linear regression on the
peak locations $\beta(L)$ of the specific heat and Binder's parameter according
to the law \begin{equation} \beta^\infty = \beta(L) + \frac{a_1}{L} +
\frac{a_2}{L^2} \end{equation} that was also used by
Baig~et~al.~\cite{Baig2004}, and fitted the inverse temperatures.
The statistical errors of the constant $a_2$ turned out to be of
the same order as the value itself, therefore we set $a_2=0$ for the fits in
Table~\ref{tab:fixedresults}, intentionally neglecting the contribution
$\mathcal{O}(L^{-2})$. The best fits and the energy probability density are
shown in Figure~\ref{fig:fit:orig-fbc} and the weighted average of inverse
transition temperatures is given by:
\begin{eqnarray}
  \beta^\infty &= 0.551\,38(5) \qquad & \textrm{original model, fixed bc}.
\end{eqnarray}

This estimate of the inverse transition temperature is thus in excellent
agreement with the other results obtained here from multicanonical simulations
with periodic boundary conditions for the gonihedric Ising model
$\beta^\infty = 0.551\,332(8)$ and the dual model $\beta^\infty = 0.551\,43(7)$.

Direct comparison to Ref.~\cite{Baig2004} shows that while inverse transition
temperatures are reproduced, the extremal values of observables are not. The
following observations may help to clarify the deviations.  The authors of
Ref.~\cite{Baig2004} simulated the system by applying periodic boundary
conditions and fixing one plane parallel to the $xy$-, $yz$- and $zx$-planes
each. If their simulation box consisted of a total number of $\hat{L}^3$ spins,
they simulated $(\hat{L} - 1)^3$ \emph{free} spins. Thus our data with lattices
of linear length $L$ has to be compared to their data {with} $\hat{L} = L + 1$.
Also, their specific heat and magnetic susceptibility $\hat{\chi} = \beta
\hat{L}^{-d} \left( \langle M^2 \rangle - \langle M \rangle^2 \right)$ have to
be multiplied by a factor of $(L+1)^3/L^3$ to be comparable with our
normalization, since these quantities are proportional to the inverse system
volume. Here, $M=\sum_i\sigma_i$ is the total magnetization, which for fixed 
boundary conditions is a well defined order parameter.

Binder's energy parameter has no explicit volume-dependence by design, but it
is sensitive to offsets in the energy, which cancel in the specific heat. Our
values of the Binder parameter minima differ significantly
from~\cite{Baig2004}. However, if we shift our measured energies $E$ to get
$\hat{E} = E - 1.5\hat{L}^2 = E - 1.5(L+1)^2$ and calculate Binder's
parameter~\eref{eq:binder} with $\hat{E}$ instead of $E$, our measurements
reproduce {those} of~\cite{Baig2004} very well. The energy $E$ of the system can
be written in terms of the number of plaquettes with an even or odd number of
parallel aligned spins, $n_+$ or
$n_-$, \begin{eqnarray} E(\beta) = - \frac{1}{2}\left( n_+(\beta) - n_-(\beta)
  \right).  \end{eqnarray} Since we measure the same cumulant values for
shifted energies $\hat{E}$, in Ref.~\cite{Baig2004} an
additional number of $\hat{n}_+ = n_+ + 3\hat{L}^2$ plaquettes contribute to
the system's energy because energetic contributions from the
fixed planes, where all spins are aligned, were included.

\begin{table}[htpb] \tablesize \centering 
  \caption{Values for comparison with Ref.~\cite{Baig2004}, with the hat
  denoting the observables as calculated by Baig~et~al.  Linear lattice lengths
  are $\hat{L} = L + 1$, with $L$ being the number of free spins in each
  direction. The specific heat $\hat{C}_V^{\rm max} = L^3/(L+1)^3 C_V^{\rm
  max}$, Binder's parameter $\hat{B}_{\rm min} = 1.0 -
  \langle\hat{E}^4\rangle/3\langle\hat{E}^2\rangle^2$ with $\hat{E} = E -
  1.5\hat{L}^2$. The inverse temperatures $\beta$ are the same for their data
  and ours. Magnetic susceptibilities $\hat{\chi}^{\rm max} = L^3/(L+1)^3 \chi^{\rm max}$ are
listed as well.} \vspace{3ex} 
\begin{tabular}{ll*{6}{r}} 
  \toprule
  \multicolumn{1}{c}{$L$} & \multicolumn{1}{c}{$\hat{L}$} &
  \multicolumn{1}{c}{$\beta^{\hat{C}_V^{\rm max}}$} &
  \multicolumn{1}{c}{$\hat{C}_V^{\rm max}$} & 
  \multicolumn{1}{c}{$\beta^{\hat{B}^{\rm min}}$} & 
  \multicolumn{1}{c}{$\hat{B}^{\rm min}$} & \multicolumn{1}{c}{$\beta^{\hat{\chi}^{\rm max}}$} &
  \multicolumn{1}{c}{$\hat{\chi}^{\rm max}$}\\ \midrule
9  & 10 & 0.45794(13)          & 5.56(6)              & 0.45527(14)          &
0.63964(29)        & 0.45690(14)          & 6.86(9)      \\ 11 & 12 &
0.47465(14)          & 12.18(11)            & 0.47338(14)          &
0.63555(30)        & 0.47438(14)          & 16.37(15)    \\ 13 & 14 &
0.48629(9)           & 25.27(24)            & 0.48559(9)           & 0.6282(5)
& 0.48621(9)           & 36.5(4)      \\ 14 & 15 & 0.49086(8)           &
35.5(4)              & 0.49032(9)           & 0.6234(6)          & 0.49083(9)
& 52.9(7)      \\ 17 & 18 & 0.50126(7)           & 78.0(6)              &
0.50096(7)           & 0.6133(6)          & 0.50125(7)           & 126.2(1.2)
\\ 19 & 20 & 0.50648(8)           & 121.7(1.0)           & 0.50626(8)
& 0.6062(7)          & 0.50648(8)           & 206.9(1.9)    \\ \bottomrule
\end{tabular} \label{tab:fixed_compare} \end{table} 

For direct comparison, the resulting quantities are listed in
Table~\ref{tab:fixed_compare} after applying all corrections, showing that our
data is then in very good agreement with Ref.~\cite{Baig2004}. For completeness
we include here also our data for the magnetic susceptibility $\hat{\chi}$.
The deviation from the fitting results in Ref.~\cite{Baig2004} simply stems
from the fact that our simulations are performed with the multicanonical 
method that allows a finite-size scaling analysis with more and significantly
larger lattice sizes. 

{
  The interface tension as a function of the linear lattice size~\eref{eq:interface-tension} is extracted and its infinite-volume limit yields a value of $\sigma = 0.0281(7)$ where we allowed corrections of order ${\cal O}(L^{-2})$ in the fits. Note that this value is about four times smaller than that for the same model with periodic boundary conditions.
}

\section{Summary} 

We simulated the plaquette gonihedric Ising model and its dual to shed some
light on discrepancies in the recent literature on the reported value(s) of the
first-order phase transition temperature.  High-precision results from
multicanonical simulations forced us to review the traditional scaling ansatz
for first-order finite-size corrections by taking the exponential low-temperature 
phase degeneracy of the model into account. The leading correction
in such circumstances is then no longer proportional to the inverse volume of
the system, ${\cal O}\left( L^{-3} \right)$, but is rather ${\cal O}\left(
L^{-2} \right)$. With this finite-size scaling ansatz, our simulations with
periodic boundary conditions produced consistent results for both the original
formulation of the model as well as its dual representation.  Since our results
also differed {from} earlier simulations {using} fixed boundary conditions, where
the leading corrections are now ${\cal O}\left( L^{-1} \right)$, we carried out
multicanonical simulations of the gonihedric Ising model with these boundary
conditions too. The resulting inverse transition temperature was fully
consistent with the value found using periodic boundary conditions when larger
lattices were included, and hopefully settle once and for all enduring
inconsistencies.
Interestingly, we do find different values for the latent heat for different
boundary conditions, $\Delta\hat{e} = 0.694(4)$ in the case of fixed
boundary conditions compared to $\Delta\hat{e} = 0.850968(18)$ for periodic
boundaries. That the latent heat may be dependent on the boundary conditions has
been observed earlier for the $q$-states Potts model~\cite{fssfbc}.

The main resulting physical
quantities that characterize the first-order phase transition are summarized in
Table~\ref{tab:infinite_system_results} for the different models and boundary
conditions in our simulations.
We find an overall consistent value for the inverse transition temperature of 
\begin{equation}
    \beta^\infty = \betainfall
\end{equation}
and we measure the interface tension of the original model and its dual for the 
first time.
{
We find values of $\sigma = 0.12037(18)$ 
and  $\sigma = 0.1214(13)$ for the original and dual model with periodic boundary conditions, respectively. The interface tension of the original model with fixed boundary conditions is found to be much smaller, $\sigma = 0.0281(7)$.}

\begin{table}[tpb] 
  \tablesize 
  \centering 
  \caption{
    Overview of resulting quantities of the infinite systems. } \vspace{3ex}
    \begin{tabular}{*{7}{l}} 
    \toprule \multicolumn{1}{c}{model} &
    \multicolumn{1}{c}{bc} & \multicolumn{1}{c}{$\beta^\infty$} &
    \multicolumn{1}{c}{$\hat{e}_{\rm o}$} & \multicolumn{1}{c}{$\hat{e}_{\rm d}$} & \multicolumn{1}{c}{$\Delta\hat{e}$} &
    \multicolumn{1}{c}{$\hat\sigma$}\\ 
    \midrule
    original~\eref{eq:ham:gonikappa0}  & periodic & 0.551332(8) & $-1.468364(4)$ & $-0.617396(11)$ & 0.850968(18) & {0.12037(18)} \\
    dual~\eref{eq:ham:dual}            & periodic & 0.55143(7)  & $-1.37644(21)$ & $-0.88227(19)$  & 0.49402(26)  & 0.1214(13)\\ 
    &          & \multicolumn{2}{l}{[$\beta^\infty_{\rm dual} = 1.31328(12) $]}\\ 
    original~\eref{eq:ham:gonikappa0} & fixed    &  0.55138(5)  & $-1.4782(27) $ &$ -0.790(4)  $ & 0.694(4)  & 0.0281(7) \\ 
    \bottomrule
    \end{tabular} \label{tab:infinite_system_results} \end{table} 
{
Any model with an exponentially degenerate low-temperature phase will display the modified scaling at a 
first-order phase transition we have delineated for the {three-dimensional} gonihedric model and its dual here.
Apart from higher-dimensional variants of the gonihedric model or its dual, there are numerous other models in which the scenario could be realized. 
Examples range from ANNNI models \cite{selke} to topological {``orbital''} models in the context of quantum computing \cite{nussinov} which all share an extensive ground-state degeneracy. Among the orbital models for transition metal compounds, a particularly promising candidate is the three-dimensional classical compass or $t_{2g}$ orbital model   where a highly degenerate ground state is well known and the signature of a first-order transition into the disordered phase has recently been found numerically \cite{compass4}.}

{
	Other systems, such as the {three-dimensional} Ising antiferromagnet on {an} FCC lattice, have an exponentially
	degenerate number of ground states but a small number (6 in the case of the {FCC} Ising antiferromagnet) of true low-temperature phases. Nonetheless, they do possess
an exponentially degenerate  number of low-energy excitations so, depending on the nature of the growth of energy barriers with system size, an {\it effective} modified scaling could still be seen at a first-order transition for the lattice sizes accessible in typical simulations. The crossover to the true asymptotic (standard) scaling would then only appear for very large lattices.
Indeed, previous simulations appear to have found non-standard scaling for the first-order transition in 
the {three-dimensional} Ising antiferromagnet on {an} 
FCC lattice \cite{beath_ryan}.}

\section*{Acknowledgment}
This work was supported by the Deutsche Forschungsgemeinschaft (DFG)
through the Collaborative Research Centre SFB/TRR 102 (project B04) and by
the Deutsch-Franz\"osische Hochschule (DFH-UFA) under Grant No.\ CDFA-02-07.

\end{document}